\documentclass[aps,nofootinbib,twocolumn,longbibliography]{revtex4}

\usepackage[utf8]{inputenc}
\usepackage{amsmath,amsfonts,amssymb}
\usepackage{graphicx}
\usepackage{float}
\usepackage[final]{pdfpages}
\usepackage[normalem]{ulem}
\usepackage{comment}
\usepackage{hyperref}

\hypersetup{
	colorlinks=true,
	linkcolor={red!90!black},
	citecolor={blue!90!black},
    urlcolor={blue!80!black}
}

\newcommand{\gdot}{\dot{\gamma}}

\begin{document}

\title{A constitutive model for discontinuous shear thickening in epithelial tissues}

\author{Tanmoy Ghosh}
\affiliation{Tata Institute of Fundamental Research, Hyderabad - 500046, India}
\author{Kabir Ramola}
\affiliation{Tata Institute of Fundamental Research, Hyderabad - 500046, India}
\author{Saroj Kumar Nandi}
\affiliation{Tata Institute of Fundamental Research, Hyderabad - 500046, India}

\begin{abstract}

The rheological properties of biological tissues, though fundamental to many physiological and pathological processes such as embryonic development, wound healing, and tumor progression, remain poorly understood. A recent study showed that the active vertex model of biological tissues exhibits discontinuous shear thickening (DST), where stress and viscosity suddenly increase at a critical shear rate. What is the mechanism of DST here? Is it another nontrivial feature of activity or an inherent property of the system? To address this, we show that the thermal vertex model also exhibits DST at a small but non-zero temperature $T$. Solid-like and liquid-like cells coexist at the stress jump, and the stress-controlled flow curves exhibit the characteristic S-shape. We then introduce a constitutive model for DST in epithelial tissues. As $p_0$ increases, the theory predicts DST, followed by continuous shear thickening (CST), and finally Newtonian behavior, consistent with simulations. DST begins at the jamming point, $p_0^m$, and the Newtonian behavior starts at $p_0^*$, where the yield stress vanishes. Both $p_0^*$ and the liquid-to-solid transition stress, $\sigma^*$, govern the DST-CST boundary. Furthermore, $p_0^*$ and $\sigma^*$ also depend on $T$. Increasing $T$ reduces $p_0^*$, narrows the shear-thickening regime, and eventually destroys DST when $p_0^* \leq p_0^m$. Thus, the primary ingredients of DST in tissue models are a finite yield stress in the unjammed regime and non-zero fluctuations, whose specific form is not important. The theory agrees well with our simulation data and also provides further testable predictions.

\end{abstract}
\maketitle

\section{Introduction}

As embryos develop, morphogenesis occurs, wounds heal, or cancer progresses, tissues experience continuous mechanical stresses. In these systems, mechanical stress can be both internal and external. For example, during the formation of the Drosophila wing, the contraction of the hinge induces cell flows in the wing blade, leading to a change in the shape of the wing and a reorientation of the planar polarity~\cite{aigouy2010}. During the formation of the palate in Zebrafish, neighboring tissues act as barriers, generating stresses in the shaping of the palate~\cite{swartz2011}. On the other hand, during carcinoma progression~\cite{cheung2013,labernadie2017,palamidessi2019} and wound healing~\cite{poujade2007,tetley2019}, cells generate active stresses leading to collective cellular movements. The dramatic shape changes during asexual reproduction in {\it Trichoplax adherens} arise from internal, motility-driven stress~\cite{prakash2021}. Besides, the time and length scales of physiological tissue rheology can be diverse. During development, the strain rate is typically of the order of $10^{-2}$\%$s^{-1}$, whereas during normal body functions in adult animals, it can be of the order of $10-100 \% s^{-1}$. The deformation time scale can also vary. For some organs, such as the lungs and heart, it can last for seconds, whereas for others, such as the skin, intestine, and bladder, it can last for minutes~\cite{blanchard2009,duque2024,papafilippou2025}. How the epithelium, a confluent monolayer of cells serving as a protective layer for most organs, responds to such diverse stresses and varying time scales is crucial for both health and disease~\cite{bonfanti2022}. 

Thus, the rheological properties of epithelial tissues have great significance~\cite{kayal2025}. Past research has shown that these systems exhibit glassy properties, residing near the interface between a solid and a liquid~\cite{park2015,bi2015,bi2016,sadhukhan2024}. In addition, they have unusual rheological properties: unlike ordinary granular materials, the elastic modulus, $G$, vanishes above the jamming point, but the yield stress remains finite, $\sigma_y$~\cite{huang2022,fielding2023,berthier2025}. Furthermore, biological tissues exhibit several forms of active fluctuations~\cite{bi2016,tjhung2017,li2025,fernandez2017,czajkowski2019,sadhukhan2021} and are often subjected to mechanical stresses from other tissues. For example, during palate formation in Zebrafish, cells must move through the extracellular space, and the presence of other tissues influences the palate shape~\cite{swartz2011}. Cancer cells generate large active forces via their actomyosin contractility while being subjected to pressure from their rapid growth, surrounding tissues, and the extracellular matrix~\cite{clark2015,labernadie2017,cheung2013}. Thus, the rheological properties in the presence of active fluctuations have fundamental importance. In a recent work, Hertaeg {\it et al} have shown~\cite{hertaeg2024} that a confluent active vertex model of epithelial tissues under external shear shows Discontinuous Shear Thickening (DST): a phenomenon where the viscosity abruptly and substantially increases at a critical shear rate, causing the material to transition from a low-viscosity flowing state to a high-viscosity or shear-jammed state.

Discontinuous shear thickening in dense granular suspensions has been extensively investigated. Early experiments established the existence of sharp viscosity jumps, stress-controlled hysteresis, and re-entrant jamming transitions~\cite{brown2009dynamic,fall2008shear,fall2010shear}. Subsequent works clarified the roles of dilation, confinement, and stress transmission in governing the transition~\cite{brown2012role}, as well as the central importance of stress-controlled rheology and frictional interactions in dense suspensions~\cite{wagner2009shear,brown2014shear,guazzelli2018rheology,baumgarten2019}. Systematic studies further unified suspension and granular rheology across viscous and inertial regimes~\cite{boyer2011unifying,trulsson2012transition,fernandez2013,mari2014shear}, highlighting the proximity to jamming as the organizing principle~\cite{boyer2011unifying,andreotti2012shear}. A major conceptual advance came with the constitutive model of Wyart and Cates~\cite{wyart2014}, which attributes DST to a stress-induced transition from frictionless to frictional rheology governed by the fraction of frictional contacts. In this picture, at low applied stress, lubrication layers mediate particle motion, allowing sliding and rotation, leading to an effectively frictionless response. Above a critical stress, these lubrication layers break down, and frictional contacts proliferate, constraining rotational and sliding degrees of freedom and producing an abrupt increase in viscosity. Later theoretical developments have expanded this framework in several directions, from scaling theories near jamming to universal scaling frameworks inspired by renormalization that interpolate between frictionless and frictional critical points~\cite{malbranche2022scaling,ramaswamy2025universal}. Microscopic approaches have resolved the structure of force networks and contact correlations both in real space and in force space~\cite{royer2016rheological,thomas2018microscopic}, and constitutive models have been developed that explicitly incorporate frictional contact populations and stress-dependent rheology~\cite{mari2014shear}. Alternative perspectives have emphasized that hydrodynamic interactions alone may generate similar rheological signatures in systems of rough particles~\cite{jamali2019}, underscoring that while the phenomenology of DST is robust, the microscopic mechanisms may vary across systems.

In comparison, tissue models are fundamentally different. First, they are confluent, meaning that cells cover the entire space, leaving no intercellular spaces to accommodate liquid. Second, the notion of frictionless and frictional jamming rheology in particulate systems~\cite{boyer2011,wyart2014,jamali2019} does not appear for tissue models. And third, they differ microscopically due to many-body interactions. These differences imply that the mechanism governing DST in the active vertex model under external shear must differ from that in dense granular suspensions. Besides, activity can have nontrivial effects in confluent models, such as fluidization and unjamming~\cite{bi2016,activematterreview,kim2020,mitchel2020}, flocking~\cite{giavazzi2018}, clustering and phase separation~\cite{rozman2024,graham2024,sahu2020}, etc. Therefore, it is not clear whether DST is a characteristic of the model or comes from the properties of activity that can have further intricacy in confluent models~\cite{sadhukhan2024}.

In this paper, we first show that a thermal vertex model also exhibits DST at a finite but non-zero temperature $T$. In tissue models, $T$ should be viewed as the zero-persistence-time limit of the active forces. We systematically explore the rheological properties across a range of parameter regimes. We find that the system shows only shear thinning at $T=0$; thus, the presence of some fluctuations is essential for shear thickening. Our results show that DST is a property of confluent systems, but it requires fluctuations—either active, as in Ref.~\cite{hertaeg2024}, or thermal, as here. To further elucidate the basic principles of DST in tissue models, we develop a constitutive model for DST in epithelial tissues. Similar to the Wyart-Cates model, the fraction of solid-like and liquid-like cells also governs DST; however, the mechanism of DST here is fundamentally different. We organize the rest of the paper as follows: We introduce the model and provide the simulation details in Sec.~\ref{model}. We then present the results of this work. We first show, in Sec.~\ref{noST_zeroT}, that there is no shear thickening at zero $T$, and, in Sec.~\ref{DST_finiteT}, that the system shows DST at finite $T$. We present our constitutive model in Sec.~\ref{theory} and the predictions in Sec.~\ref{predictions}. We conclude the paper in Sec.~\ref{disc} by discussing the implications of our results.

\section{Model and Simulation Details}
\label{model}
We consider the two-dimensional vertex model of confluent tissue monolayers for our simulations. The model represents cells as polygons with preferred area, $A_0$, and preferred perimeter, $P_0$. The energy function~\cite{honda1980,marder1987,farhadifar2007,fletcher2014,activematterreview} governing the properties of the system is
\begin{equation}
H =  \sum_{i=1}^{N} \left[\Lambda_A(A_i - A_0)^2 + \Lambda_P  (P_i - P_0)^2 \right],
\label{eq1}
\end{equation}
where, $N$ is the total number of cells in the tissue, $A_i$ and $P_i$ are the area and perimeter of the $i$th cell. $\Lambda_A$ and $\Lambda_P$ are the area and perimeter moduli, respectively. The first term in Eq.~(\ref{eq1}) represents the incompressibility of cell cytoplasm. The second term represents the intercellular interactions and properties of the cellular cortex~\cite{sadhukhan2024,activematterreview,sadhukhan2022}. Since the target perimeter plays a crucial role in governing the properties of this system, let us write the energy function, Eq.~(\ref{eq1}), in terms of a non-dimensional target perimeter, $p_0=P_0/\sqrt{A_0}$. We divide Eq.~(\ref{eq1}) by $A_0^2$ and write $\mathcal{H}=H/A_0^2$ as
\begin{equation}
\label{normalizedH}
\mathcal{H} =  \sum_{i=1}^{N} \left[\lambda_A(a_i - 1)^2 + \lambda_P  (p_i - p_0)^2 \right],
\end{equation}
where $a_i = A_i / A_0$, $\lambda_A=\Lambda_A$, $\lambda_P = \Lambda_P / A_0$, and $p_i = P_i / \sqrt{A_0}$. The system shows a rigidity transition at $p_0\simeq 3.81$, separating a solid-like regime below and a liquid-like regime above this value~\cite{park2015,bi2015,bi2016}.

We apply shear in the $x$-direction, with a gradient in the $y$-direction. Following Refs.~\cite{ikeda2012,hertaeg2024}, we have used Lees-Edwards periodic boundary conditions (see Supplementary Material (SM), Sec. S1 for more details). As discussed in the introduction, contrary to Hertaeg {\it et al}~\cite{hertaeg2024}, we use thermal noise rather than active noise as the source of fluctuations. Within the vertex model, the vertices are the degrees of freedom. We evolve the vertices via overdamped Langevin dynamics; the equation of motion for the $i$th vertex position, $\vec{r}_i$, is
\begin{equation}\label{eqofmotion}
\xi\left(\frac{\partial \vec{r}_i}{\partial t} - \dot{\gamma} y_i \hat{e}_x \right)= -\nabla_i \mathcal{H}  + \vec{f_i},
\end{equation}
where $\xi$ is the damping coefficient that we set to unity, $\nabla_i$ is the derivative with respect to $\vec{r}_i$, $\dot{\gamma}$ denotes the shear rate and $\vec{f_i}$ represents the thermal fluctuations that obey $\langle \vec{f}_i\rangle=0$ and $\langle \vec{f}_i(t) \vec{f}_j(t_0) \rangle = 2 T  \delta_{ij} \delta(t - t_0)$. The first term on the right-hand side of Eq.~(\ref{eqofmotion}) gives the force coming from the energy function.

The stress tensor~\cite{chiou2012, tong2023,yang2017} for each cell $i$, denoted by $\boldsymbol{\sigma}_i$, is defined as $\boldsymbol{\sigma}_i = -\Pi_i \hat{\mathbf{I}} + \frac{1}{2 A_i} \sum_{e \in i} \mathbf{T}_e \otimes \mathbf{l}_e$, where the summation runs over all junctions $e$ belonging to cell \(i\). Here, the hydrostatic pressure inside the cell is $\Pi_i = -\frac{\partial H}{\partial A_i} = - 2 \Lambda_A (A_i - A_0)$, and $\hat{\mathbf{I}}$ denotes the identity tensor. The tension along a junction $e$ is $\mathbf{T}_e = \frac{\partial H}{\partial \mathbf{l}_e} = 2 \Lambda_P (P_i - P_0)\, \frac{\mathbf{l}_e}{|\mathbf{l}_e|}$, where $\mathbf{l}_e$ is the vector joining the two vertices defining the junction. To characterize the overall mechanical response of the system, the average stress tensor $\boldsymbol{\sigma}$ is computed as a weighted sum over all cells, $\boldsymbol{\sigma} = \sum_i w_i \boldsymbol{\sigma}_i$, with weights $w_i = \frac{A_i}{\sum_i A_i}$ proportional to cell areas. In this paper, we focus on the shear stress component $\sigma_{xy}$ and call it $\sigma$. We have investigated both rate-controlled and stress-controlled behaviors (see SM Sec. S6 for more details). For the latter, we have used a feedback mechanism, as described in Refs.~\cite{vezirov2015, cabriolu2019, chaudhuri2025}, to maintain the macroscopic stress at a target value $\sigma_0$. We then calculate the shear rate, $\gdot$, that provides the desired stress via
\begin{equation}
	\frac{d \dot \gamma}{d t} = B [\sigma_0 - \sigma(t)],
\end{equation}
where $B$ is a damping parameter, we have used $B=1$ in our simulations.

\begin{figure*}
	\includegraphics[width=17cm]{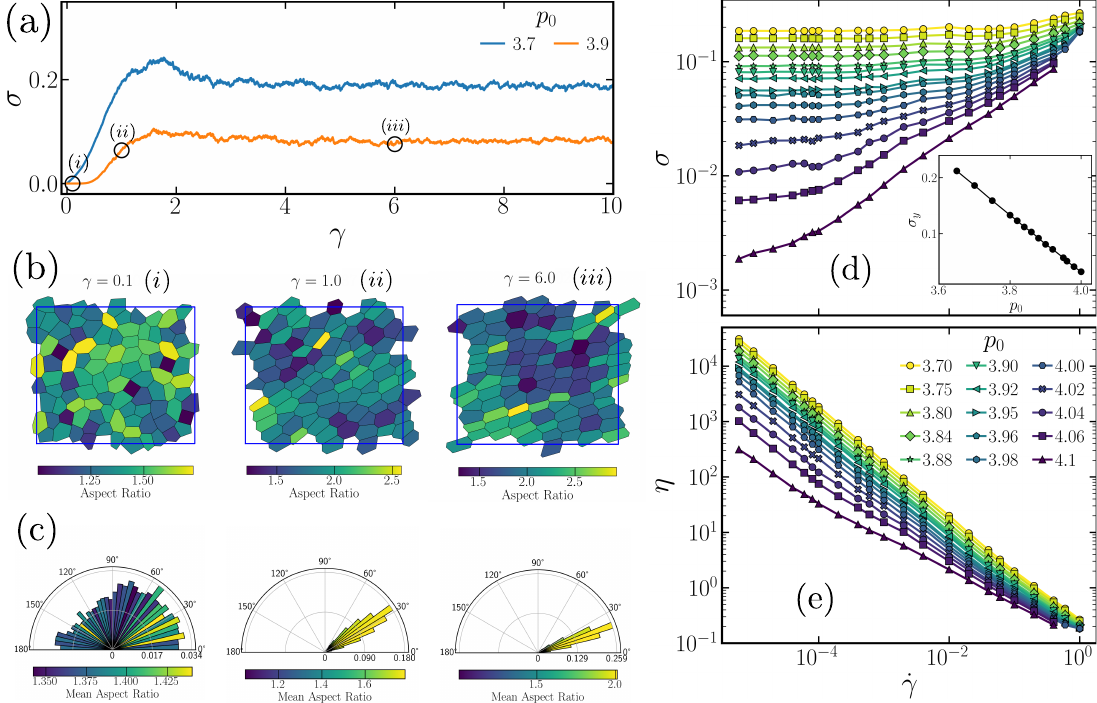}
	\caption{Rheological properties at $T=0$. (a) Stress-strain curves for two representative target shape indices, $p_0 = 3.7$ and $p_0 = 3.9$. When the strain $\gamma$ is small, the stress, $\sigma$, increases with $\gamma$ and then reaches a steady state at large $\gamma$. (b) Representative snapshots at three selected strain values for $p_0 = 3.9$ [marked in (a)]; the colors encode the local aspect ratio. (c) Histogram of the orientation of the principal axes with respect to the flow direction; the color scale represents the magnitude of the average aspect ratio within each angular bin. When $\gamma$ is very small, the cells are less elongated and randomly oriented. As $\gamma$ increases, the cells elongate and orient along the shear direction. (d) Steady state flow curves, $\sigma$ as a function of $\gdot$, for different $p_0$. We can fit each curve to the Herschel-Bulkley form to extract the yield stress, $\sigma_y$, which decreases with increasing $p_0$ (inset). (e) Shear viscosity, $\eta=\sigma/\gdot$, for the same flow curves as in panel (d). The system only shows shear thinning at $T=0$. We have used $N=100$ for these simulations.}
	\label{zeroT}
\end{figure*}

A two-dimensional monodisperse system, where $A_0$ is the same for all cells, forms a periodic hexagonal ground state. To avoid this crystallization, we have used a $50:50$ binary mixture with $A_{0\alpha} = 0.8$ and $A_{0\beta} = 1.2$. We have used $\lambda_A=1.0$ and $\lambda_P=1.0$ in our simulations. The topological T1 transition, in which an edge flips perpendicular to the existing edge when it becomes smaller than a specific value, plays a crucial role in the vertex model. We have chosen the threshold for this transition as $\ell=0.04$. We prepare the system as follows: we first equilibrate it at a relatively high temperature, $T=0.1$, then perform an instantaneous quench to $T=0$, and finally allow it to evolve to minimize its energy. We employ shear on this well-minimized state. We use $\sqrt{A_0}$, where $A_0$ is the average target area, as the unit of length and $1/(\xi \lambda_A A_0)$ as the unit of time. We take the smallest time step, $dt=0.01$. For the rate-controlled simulations, we used $N = 100$, and for the stress-controlled simulations, $N = 400$; we have checked that there are no finite-size effects for the rate-controlled simulations (SM Fig. S5). For the configuration plots, we used $N = 100$ for better visualization. For the steady state data, we have excluded the initial transient behavior from the analysis. We now present the main results of this work.

%% ======================================================

\section{Results}

We first present simulation results that clarify the main ingredients of discontinuous shear thickening (DST), then describe the constitutive model capturing DST, and finally the theoretical predictions.

\subsection{No shear thickening at zero temperature}
\label{noST_zeroT}
The jamming properties of the vertex model have distinctive features compared to those in particulate models. In the absence of thermal fluctuations, when temperature $T=0$, the shear modulus, $G$, disappears around $p_0 = p_0^m \simeq 3.81$, indicating a transition from a solid-like regime for $p_0<p_0^m$ to a liquid-like behavior for $p_0 > p_0^m$. Note that although $p_0^m$ is around $3.81$, its exact value depends on the preparation protocol~\cite{pinaki2010}; we will obtain its precise value for our system later. The liquid-like regime for $p_0>p_0^m$ is crucially different from ordinary liquids as the yield stress, $\sigma_y$, remains non-zero and vanishes at an even larger value of $p_0=p_0^*\simeq 4.06$~\cite{huang2022,fielding2023}. Thus, the system is a solid with non-zero $G$ and $\sigma_y$ in the regime $p_0<p_0^m$. By contrast, $G=0$ but $\sigma_y$ remains non-zero when $p_0^m<p_0<p_0^*$ and leads to distinct transient and steady state shear properties in this regime. On the other hand, one fundamental aspect of the vertex model is that the system properties are intimately related to cell shapes~\cite{atia2018,sadhukhan2022}. As shear increases, the cells become more elongated and orient along the shear direction, increasing their perimeters and enhancing stress on individual cells.

Figure~\ref{zeroT}(a) shows the typical behavior of $\sigma(\gamma)$ with applied shear, $\gamma$, for $p_0=3.70$ and $p_0=3.90$, one below $p_0^m$ and the other above. The applied stress, $\sigma(\gamma)$ increases as $\gamma$ increases, eventually the solid yields beyond a critical $\gamma$, and $\sigma(\gamma)$ reaches a steady state value, $\sigma$ (Fig.~\ref{zeroT}(a)). The cell shape variation with increasing shear at $T=0$ is qualitatively similar for both $p_0$ values. Figure~\ref{zeroT}(b) shows typical snapshots of the cells in our system at three values of the shear [as marked in (a)], clearly showing the cell elongation and major axis orientation with increasing shear. The bar-graph in Fig.~\ref{zeroT}(c) shows the magnitude of aspect ratio (AR, which quantifies cell elongation, see SM Sec. S4 for the definition) by the length of the bars, and the angle represents the major axis orientation with respect to the shear direction at three different values of shear. When $\gamma$ is small (first plot in Fig.~\ref{zeroT}(c)), cell elongation, and hence AR, is small. In addition, the major axes of the individual cells are randomly oriented. As $\gamma$ increases (second plot in Fig.~\ref{zeroT}(c)), AR increases as the cells now orient towards the shear direction. Finally, when $\gamma$ becomes much larger, where the system reaches steady state (third plot in Fig.~\ref{zeroT}(c)), the cells are even more elongated and oriented along the shear direction. As deformed cells have higher perimeters, cells with larger AR will have greater stress (see Sec.~\ref{model}). Thus, snapshots of the experimental system can provide insights into cell stress and other rheological properties of the tissue (see SM Secs. S5 and S6 for additional measures).

We now study the flow properties of the system in the steady state after the solid has yielded. The steady state value $\sigma$ depends on the shear rate, $\gdot$ (see SM Fig. S8). Figure~\ref{zeroT}(d) shows the flow curves of $\sigma$ as a function of $\gdot$ for several values of $p_0$ in the range $3.70\leq p_0\leq 4.10$. We fit each curve with a Herschel-Bulkley form~\cite{herschelBulkley,bonn2017,nicolas2018}, $\sigma=\sigma_y+B\gdot^n$, using $\sigma_y$, $B$ and $n$ as fitting parameters. From the fit, we obtain $\sigma_y(p_0)$. The inset of Fig.~\ref{zeroT}(d) shows $\sigma_y$ as a function of $p_0$: $\sigma_y$ decreases with increasing $p_0$ and a fit with a linear form shows that it will reach zero at $p_0^*\simeq4.06$. This value of $p_0^*$ is consistent with existing results~\cite{huang2022}. The system for $p_0>4.06$ behaves as an ordinary liquid with both $G$ and $\sigma_y$ being zero. Note that $\sigma_y$ smoothly varies around $p_0=p_0^m$, the jamming point associated with the vanishing of $G$. We can define viscosity $\eta$ in the flowing phase via $\eta=\sigma/\gdot$. Figure~\ref{zeroT}(e) shows $\eta$ with $\gdot$ for the data corresponding to the flow curves presented in Fig.~\ref{zeroT}(d): $\eta$ monotonically decreases with increasing $\gdot$. Thus, the system shows {\it only} shear thinning, and no signature of shear thickening, designated with increasing $\eta$, at $T=0$. We will next show that this behavior changes dramatically in the presence of a small but nonzero $T$.

\begin{figure*}
	\centering
	\includegraphics[width=15cm]{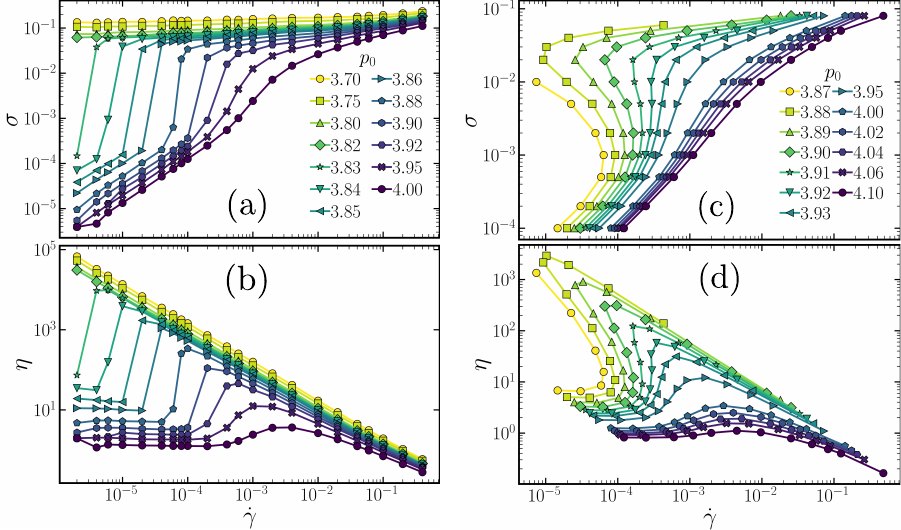}
	\caption{Discontinuous shear thickening at finite $T$. (a) Flow curves for different values of $p_0$ at $T = 2 \times 10^{-4}$ for a system size $N = 100$. (b) Shear viscosity $\eta$ corresponding to the flow curves shown in panel (a). (c) S-shaped flow curves obtained under stress-controlled conditions at $T = 2 \times 10^{-4}$ for $N = 400$. (d) Shear viscosity corresponding to the flow curves shown in panel (c). The sudden jump in both $\sigma$ and $\eta$ and the S-shaped flow curves are key characteristics of DST.
	}
	\label{flow}
\end{figure*}

\subsection{Discontinuous shear thickening at finite $T$}
\label{DST_finiteT}
The rheological behavior at a small finite $T$ exhibits remarkably rich physics, as we will show below. In the absence of shear, the thermal fluctuations govern the dynamics with a characteristic relaxation time, $\tau$, that varies with $p_0$~\cite{bi2016,sadhukhan2024}. In the $(p_0,T)$ plane, the system shows a glass transition line, $p_{0,g}(T)$ where $\tau$ attains a certain value, say $\tau=10^6$~\cite{bi2016,sadhukhan2021}. Although the physics of glass transition and that of jamming are different, the glass critical point coincides with $p_0^m$ as $T\to 0$, and the dynamics is governed by the jamming critical point $p_0^m$~\cite{huang2022,bi2015,bi2016}. At small $T$, $\tau$ is infinite for $p_0\leq p_0^m$ and finite for $p_0>p_0^m$; $\tau$ goes as $1/\sqrt{T}$ in the latter regime~\cite{berthier2009,berthier2009b,puneet2026} and at constant $T$, $\tau$ decreases as $p_0$ increases. 
When we impose shear with a rate $\gdot$, the competition between the two time scales, $\tau$ coming from the thermal fluctuations and $\gdot^{-1}$ coming from the external shear, governs the dynamical behaviors. 

To investigate the finite $T$ rheology, we impose a small and constant $T=2\times10^{-4}$, and study the flow properties in the steady state with varying $\gdot$. Figure~\ref{flow}(a) shows the flow curves, $\sigma$ as a function of $\gdot$, for different values of $p_0$. For $p_0<p_0^m$, the system shows a yield stress regime characterised by a stress plateau, similar to solid-like characteristics at $T=0$ (Fig.~\ref{zeroT}(a)). However, the nature of the curves changes dramatically for larger $p_0$ (larger than $3.82$): $\sigma$ is linearly proportional to $\gdot$ at low $\gdot$, but then shows a sudden jump beyond a critical strain rate, $\gdot_c$, and eventually follows a solid-like trend. This sudden stress jump, from a liquid-like to a solid-like branch, resembles the DST. To confirm DST in this regime, we plot the corresponding shear viscosity, $\eta=\sigma/\dot\gamma$, as a function of $\gdot$ for various $p_0$ (Fig.~\ref{flow}(b)). The low-$p_0$ curves show the characteristics of a solid: a divergent viscosity such as $\dot{\gamma} \to 0$ and shear thinning. However, for the larger-$p_0$, $\eta$ remains constant at small $\gdot$, expected for a liquid, and shows a sudden jump at $\gdot_c$ followed by shear thinning behavior, as anticipated for a fluid when $\gdot^{-1}<\tau$. This abrupt jump in both shear stress and viscosity at a critical shear rate confirms DST. At even higher $p_0$, the stress jump becomes smoother and the viscosity growth is smaller, leading to continuous shear thickening or CST. Beyond a specific value of $p_0$ ($\sim 4.06$), the shear-thickening behavior vanishes, and the system resembles a Newtonian fluid with linearly increasing $\sigma$ as $\gdot$ increases. These results demonstrate that tuning $p_0$ effectively controls the mechanical response, evolving the system from a yield-stress solid through DST regimes to a near-Newtonian flowing state.

Another defining characteristic of DST is the S-shaped, nonlinear flow curves~\cite{mari2014,wyart2014,jamali2019} observed in stress-controlled simulations. To convincingly show the presence of DST, we have also carried out the stress-controlled simulations (see SM Sec. S6 for details), and Fig.~\ref{flow}(c) shows the corresponding flow curves for the same $T$ as in the rate-controlled simulations and various values of $p_0$. The S-shape signifies metastability. For the rate-controlled simulations, we cannot capture the curves beyond the inflexion point, leading to a sharp jump in $\sigma$ (Fig.~\ref{flow}(a)). Figure~\ref{flow}(d) shows $\eta=\sigma/\gdot$ for the data in Fig.~\ref{flow}(c). The basic characteristics are similar to those in Fig.~\ref{flow}(b): a fluid regime with a constant $\eta$ at small $\gdot$ and shear thinning at large $\gdot$. In addition, the S-shape of the $\sigma$ vs $\gdot$ curves in Fig.~\ref{flow}(c) disappears at larger $p_0$, and it becomes nearly linear at even larger $p_0$: this signifies DST, followed by CST, followed by near-Newtonian behavior as we have discussed above.

\begin{figure*}[t!]
	\centering
	\includegraphics[width=17.2cm]{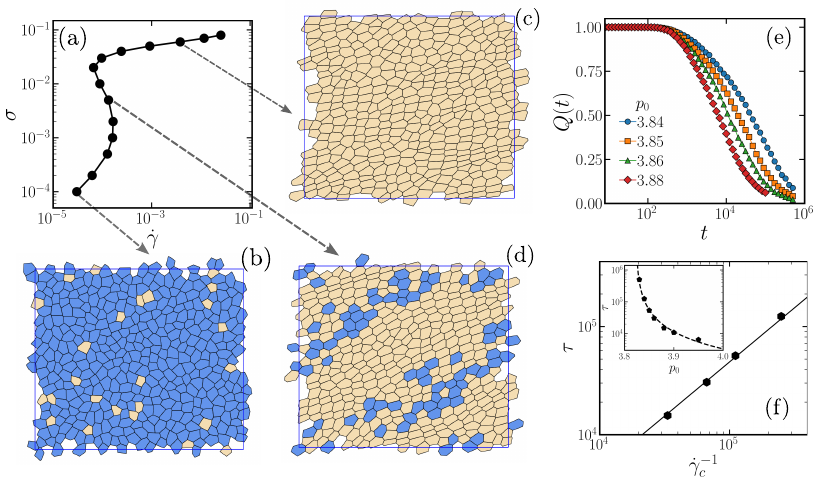}
	\caption{Fluid-to-solid transition and the relaxation dynamics. (a) Stress-controlled flow curve for $p_0 = 3.9$ at temperature $T = 2 \times 10^{-4}$. We show the physical states of the cells based on their perimeter, $p_i$, at the three points marked by the arrows as the stress transitions from the liquid-like to the solid-like regime. We mark the cell blue if $|p_i-p_0|\leq 0.02$, otherwise, yellow. (b) In the liquid-like branch, most cells have $p_i\simeq p_0$. (c) In the solid-like branch, the cells are deformed, with most having $p_i$ much higher than $p_0$. (d) In the DST regime, blue and yellow cells coexist. (e) We study the dynamics via the overlap function, $Q(t)$. At fixed $T = 2 \times 10^{-4}$, $Q(t)$ decays becomes faster as $p_0$ increases. (f) The relaxation time $\tau$ varies nearly linearly with $\dot{\gamma}_c^{-1}$, the characteristic shear rate at which $\sigma$ jumps from the liquid to solid branch. The solid line represents $\tau=1+0.47 \gdot^{-1}_c$. {\bf Inset}: $\tau$ as a function of $p_0$; symbols represent the data, and the dashed line represents $\tau = 247.34\,(p_0 - 3.823)^{-3/2}$.
		}
	\label{snapshots}
	
\end{figure*}

For a deeper understanding of the microscopic state of the system under shear, we examine the deviations of cell perimeters $p_i$ from the target perimeter $p_0$, as these reflect the stress on the cells (see Sec.~\ref{model}). Specifically, we focus on cell perimeters in the three regimes of the flow curve: the fluid, the solid, and the intermediate metastable regime as marked in Fig.~\ref{snapshots}(a) for the stress-controlled flow curve at $p_0=3.9$. Figure~\ref{snapshots}(b) shows the cell configurations in the fluid regime. When cells can satisfy the perimeter constraint of Eq.~(\ref{normalizedH}), we will have $p_i\simeq p_0$ and $\sigma$ which are quite small. To highlight the physical states of the cells, we color-code them as follows: if $|p_i-p_0|\leq 0.02$, we mark the cell blue; otherwise, we mark it yellow. We see that the cells are predominantly blue in the liquid-like regime (Fig.~\ref{snapshots}(b)) and they are mostly yellow in the solid-like regime (Fig.~\ref{snapshots}(c)). However, blue and yellow cells coexist in the intermediate phase, highlighting the coexistence of fluid-like and solid-like cells in this regime (Fig.~\ref{snapshots}(d)). Blue and yellow cells are analogs of smooth and rough contacts in the Wyart-Cates theory describing DST in dense granular suspension~\cite{wyart2014}, although the physics of the two are entirely different.

How can we understand the transition of the system from the liquid-like to the solid-like regime? As we discussed earlier, $\tau$ in the regime $p_0^m < p_0 <p_0^*$ goes as $1/\sqrt{T}$. In this sub-Arrhenius regime~\cite{sussman2018,sadhukhan2021}, $\tau$ is always finite at non-zero $T$. Moreover, the cells predominantly satisfy the perimeter constraint, resulting in very low stress in the absence of shear. As we apply shear, the cells start to elongate with increasing shear (Fig.~\ref{zeroT}(b)). When $\gdot$ is small, such that $\gdot^{-1}>\tau$, cellular elongation, and therefore stress on them, can still relax within the time scale $\tau$ and the system behaves as a liquid. By contrast, at large $\gdot$ when  $\gdot^{-1}<\tau$, cells do not have enough time to relax the effect of shear, leading to a large $\sigma$ and the system behaves as a solid. To further support this microscopic picture, we have investigated the relaxation dynamics of the system at $T=2\times 10^{-4}$ and various values of $p_0$. Figure~\ref{snapshots}(e) shows the decay of the overlap function, $Q(t)$ (see SM Sec. S2 for the definition); the decay becomes faster as $p_0$ increases. We can define the relaxation time $\tau$ when $Q(t)$ decays to $1/3$, that is, $Q(t=\tau)=1/3$. On the other hand, we also identify $\gdot_c$ from the flow curve data (as presented in Fig.~\ref{flow}). We find that $\tau\propto\gdot_c^{-1}$ (Fig.~\ref{snapshots}(f)). Thus, $\tau$ controls the liquid-like to solid-like transition as a function of $\gdot$. This mechanism also implies that temperature and activity will have similar effects as they both lead to finite $\tau$ in the absence of shear in this large $p_0$ regime~\cite{sadhukhan2021,bi2016,hertaeg2024,sadhukhan2024}. However, this transition from the fluid-like to the solid-like regime still does not explain the increase in $\eta$ that leads to shear thickening, nor, crucially, why it should be discontinuous, leading to DST. We describe below our constitutive model for the DST in this system and explain its mechanism.

\subsection{The constitutive model}
\label{theory}

Following the success of the Wyart-Cates model in predicting DST in dense granular suspensions~\cite{wyart2014}, we employ a similar strategy, interpreting the rheology of the vertex model by interpolating between two rheological branches discussed above. However, the underlying physics of DST and the interpolation mechanism differ entirely in the tissue model. In the dense suspensions, DST occurs as the frictionless rheology transitions to the frictional rheology via the formation of stress-induced contacts when the lubrication layer drains, forming an S-shaped flow curve~\cite{mari2014,wyart2014,jamali2019}. Both rheological behaviors are liquid-like, characterized by distinct critical points. By contrast, the vertex model is a confluent system with no interparticle gaps and no solvent. The rheology interpolates between a Herschel–Bulkley yield-stress fluid and a viscous liquid with divergent viscosity at a critical value of $p_0$, the jamming point $p_0^m$.

We first focus on the liquid branch. As we discussed in Sec.~\ref{DST_finiteT}, the system under shear behaves as a liquid at small $\gdot$, when $\gdot^{-1}>\tau$ (Fig.~\ref{snapshots}(f)). In this regime, $\tau$ depends on the distance from the critical point $p_0^m$. As the inset of Fig.~\ref{snapshots}(f) shows the simulation data agree well with the form $\tau\sim (p_0-p_0^m)^{-\alpha}$ with $\alpha=1.5$ and $p_0^m=3.823$. Note that $p_0^m$ is very close to the existing results~\cite{park2015,bi2016}. For a liquid, $\sigma$ must be proportional to $\gdot$, thus, $\sigma = \eta(p_0) \dot{\gamma}=K \tau(p_0)\gdot$, where we have used the Maxwell relation, $\eta=K\tau$ with $K$ being a constant. Therefore, we write the stress in the liquid branch, $\sigma=\sigma_l$, as
\begin{equation}
	\sigma_\ell = \frac{A \dot{\gamma}}{(p_0 - p_0^m)^\alpha},
	\label{liquid_eq}
\end{equation}
where $A$ is a constant. We show in the SM (Sec. S9) that the rheological data fit well with Eq.~(\ref{liquid_eq}) with $A=0.07$ in the liquid-like branch.

We next focus on the solid-like regime. When $\gdot^{-1}<\tau$, the system does not have enough time to relax the effect of shear and behaves as a yield stress solid (Fig.~\ref{snapshots}(f)). Similar to the $T=0$ data presented in Fig.~\ref{zeroT}, we can fit the steady-state shear stress in this regime via the Herschel-Bulkley relation, $\sigma = \sigma_y + B \dot\gamma^n$~\cite{herschelBulkley,bonn2017}. For $p_0 > p_0^*$, the system behaves as a liquid at all times, and the yield stress drops to zero as $\sigma_y = C (p_{0}^*-p_0)$. Thus, we can write the stress in this large $\gdot$-regime as $\sigma=\sigma_s$, 
\begin{equation}
	\sigma_s = C(p_0^* - p_0) + B \dot{\gamma}^n,
	\label{solid_eq}
\end{equation}
with the exponent $n = 0.33$ and $B$, $C$ are two constants that we obtain from the simulation data. As shown in the SM (Sec. S9), Eq.~(\ref{solid_eq}) agrees well with the simulation data using the same set of parameters: $n=0.33$, $C=0.52$, $B\simeq 0.16$, and $p_0^*=3.96$. We now interpolate between the solid-like and the liquid-like branches to obtain the DST behaviors.

\begin{figure}
	\includegraphics[width=8.6cm]{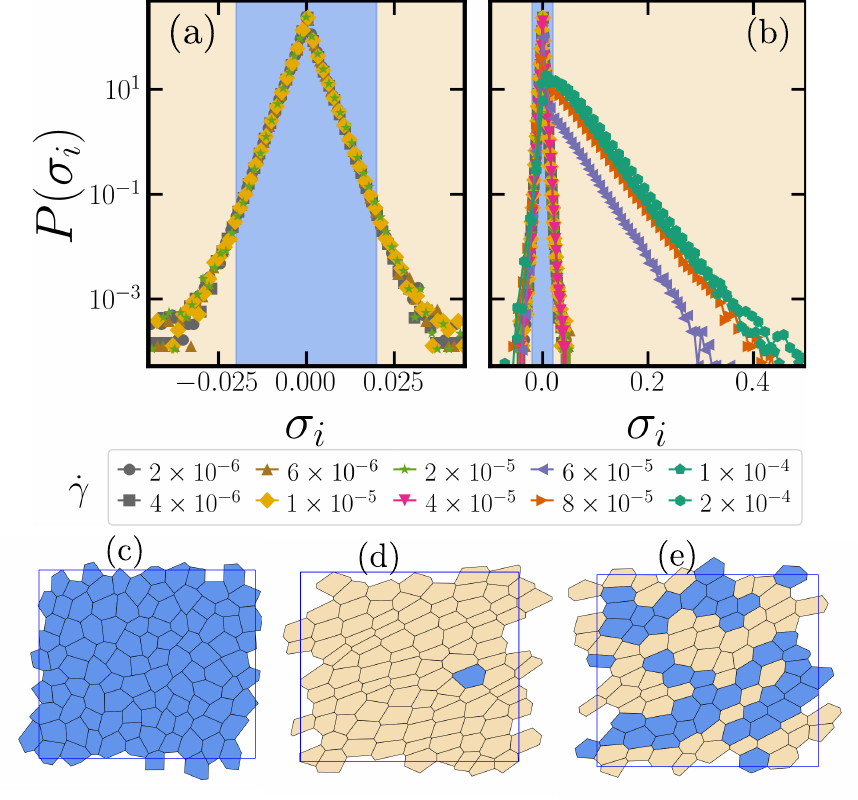}
\caption{Distribution of local stress $\sigma_i$. (a) Distribution of local cell stresses, $P(\sigma_i)$, at various strain rates $\dot{\gamma}$ in the liquid-like branch. (b) $P(\sigma_i)$ shows a broader distribution at larger $\gdot$ when the system goes to the solid-like branch. We also show $P(\sigma_i)$ in the liquid-like branch for comparison. (c)--(e) Representative snapshots illustrating three distinct rheological regimes: liquid-like, solid-like, and coexistence of liquid-like and solid-like cells. We mark the liquid-like cells blue when $|\sigma_i|\leq 0.02$, otherwise, yellow. These results are for $p_0=3.88$ and $N=100$.}
	\label{interpolation_fn}
\end{figure}

As we have discussed in Sec.~\ref{DST_finiteT}, the competition between $\tau$ and $\gdot^{-1}$ determines the cellular ability, to relax the perimeter deformation due to the applied shear, which distinguishes the liquid-like or solid-like regimes in the tissue model in the regime $p_0^m <p_0<p_0^*$. Furthermore, deviation of $p_i$ from $p_0$ determines $\sigma$ (Sec.~\ref{model}). Figure~\ref{snapshots} shows the deviation of $p_i$ from $p_0$ in various states of the system. To conclusively demonstrate that this indeed is related to applied $\sigma$, we now look at the microscopic cellular level stress, $\sigma_i$, in the system. Figure~\ref{interpolation_fn}(a) shows the distribution $P(\sigma_i)$ for a system at $p_0=3.88$ and different values of $\gdot$ in the liquid state: the distribution is quite narrow. By contrast, $P(\sigma_i)$ becomes much broader and skewed towards larger $\sigma_i$ with increasing $\gdot$, showing a significant number of cells with large values of $\sigma_i$ (Fig.~\ref{interpolation_fn}b). To show the microscopic state of the system, we mark the cells blue if the magnitude of cell stress is less than $0.02$, that is, $|\sigma_i|\leq 0.02$, and yellow otherwise. Figures \ref{interpolation_fn}(c) and (d) show representative snapshots of the system from the steady-state simulations, with the cells color-coded according to their individual stress values. Very similar to the individual perimeter distortion due to shear, nearly all cells exhibit either very small or significant stresses in the liquid-like and solid-like regimes, respectively. On the other hand, the system in the intermediate regime shows both solid-like and liquid-like cells (Fig.~\ref{interpolation_fn}e). These results show that, as in a dense granular suspension, there is a fraction of solid-like cells, $f(\sigma)$, that increases with increasing $\sigma$. We can define a cut-off value of stress, $\sigma^*$, to define solid-like or liquid-like cells; $\sigma^*$ will be a function of $p_0$ and $T$ as the system's ability to satisfy the perimeter constraint depends on both of these parameters. We will determine the actual values of $\sigma^*$ below using the simulation data.

Having demonstrated the coexistence of liquid-like and solid-like cells, we now write down the constitutive model for DST using two reference points, $p_{0l}^c$ and $p_{0s}^c$, that govern the behavior of these two rheologies. For the liquid-like behaviors, we have $\sigma_l\sim (p_0-p_{0l}^c)^{-\beta}$ and for the solid-like behavior, we have $\sigma_s\sim (p_0-p_{0s}^c)^{-\beta'}$. Here $\beta$ and $\beta'$ are two exponents. Note that we can write any functional behavior in such power law forms for generic values of $\beta$ and $\beta'$. In addition, we write the total macroscopic stress as $\sigma=(p_0-p_{0,av}^c)^{-\beta''}$, where $p_{0,av}=[1-f(\sigma)]p_{0l}^c+f(\sigma)p_{0s}^c$, interpolated between the two reference points, and $\beta''$ is another exponent. In principle, $\beta$, $\beta'$, and $\beta''$ can all be different; however, for simplicity, we assume them to be the same and equal to $\beta$. Therefore, we propose the following functional form for the overall stress as an interpolation between the liquid-like and solid-like branches
\begin{equation}\label{DSTmodel}
	\sigma = \left[ (1 - f(\sigma)) \sigma_\ell^{-1/\beta} + f(\sigma) \sigma_s^{-1/\beta} \right]^{-\beta},
\end{equation}
where we treat $\beta$ as a parameter. We will show in the next section that Eq.~(\ref{DSTmodel}) predicts DST and produces flow curves that are very similar to those in the simulations. Furthermore, the existence of DST does not require specific values of $\beta$ or the form of $f(\sigma)$ (SM, Sec. S12), as long as they satisfy certain basic conditions, demonstrating the robustness of the theory.

\begin{figure*}
	\includegraphics[width=17.2cm]{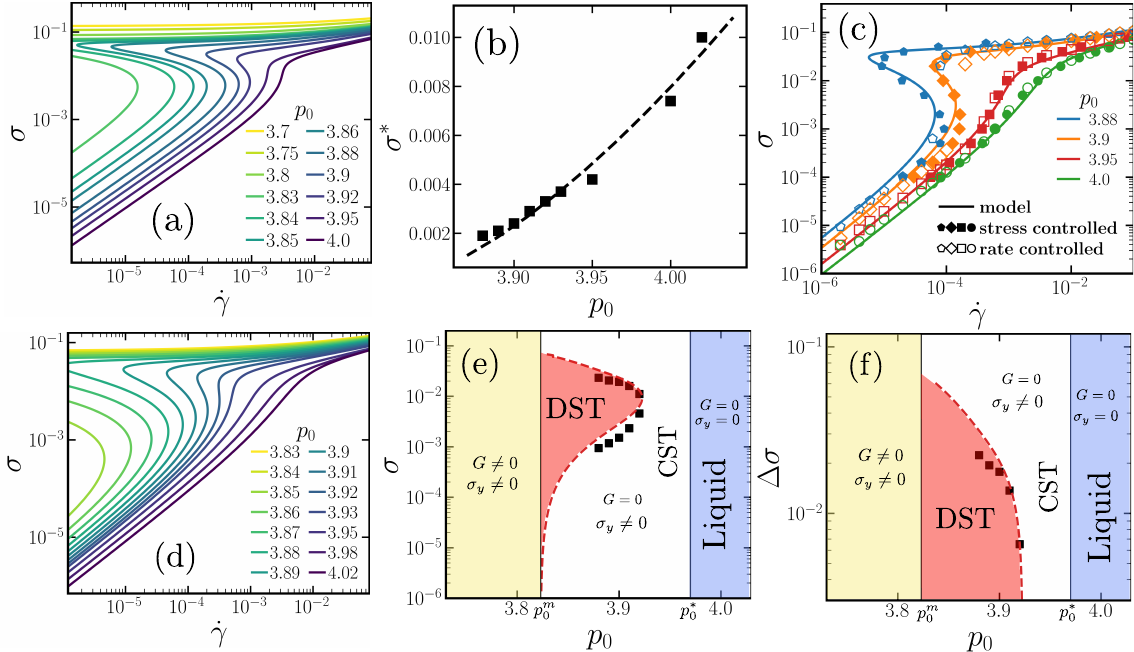}
\caption{Comparison of theoretical predictions at $T=2\times10^{-4}$ and varying $p_0$. (a) Theoretical flow curves, Eq.~(\ref{DSTmodel}), for various values of $p_0$, with $\delta = 1.0$ and a constant $\sigma^* = 0.008$. The curves show the characteristic S-shape of DST. (b) Fitting the parameters of $f(\sigma)$ with the simulation data, we find $\delta\simeq 0.6$, but $\sigma^*$ varies with $p_0$. The symbols represent the fitting values, and the line is $\sigma^*(p_0) = 0.107\,(p_0 - 3.823)^{3/2}$. (c) Comparison of the theoretical flow curves (lines) with stress-controlled (filled symbols) and rate-controlled (open symbols) simulation data for different $p_0$. (d) Theoretical flow curves using $\delta = 0.6$ and the $p_0$-dependent $\sigma^*$ from panel~(b) in the interpolating function $f(\sigma)$. The quantitative nature of the curves is different than the constant $\sigma^*$ case. (e) Comparison of the DST phase diagram between theory (dashed line) and simulations (symbols). We have indicated the different regimes: DST, then CST, and the Newtonian liquid. (f) The comparison of the stress drop, $\Delta \sigma$, plotted as a function of $p_0$ from the theory (dashed line) and simulation (symbols). The theoretical predictions agree well with the simulation data.
}
	\label{model_predictions}
\end{figure*}

\subsection{Predictions of the model and comparison with simulations}
\label{predictions}
We first show that our constitutive model predicts DST and the characteristic S-shaped flow curves. We show in the SM (Sec. S12) that the qualitative behavior does not depend on the detailed form of $f(\sigma)$, except that it should have a sigmoidal nature. One such form is the exponential: $f(\sigma)=1-\exp[-(\sigma/\sigma^*)^\delta]$, where $\sigma^*$ characterizes the transition point from the liquid-like to the solid-like behaviors and $\delta$ gives the steepness of this transition (see the SM, Sec. S12 for other forms of $f(\sigma)$). Figure~\ref{model_predictions}(a) shows the flow curves for $\delta=1$ and a constant $\sigma^*=0.008$. Similar to the simulation results, the theory also predicts shear thickening in the regime $p_0^m\leq p_0 \leq p_0^*$. The S-shaped curves appear at relatively smaller $p_0$, signifying DST. As $p_0$ increases approaching $p_0^*$, DST disappears (around $p_0=3.95$ in Fig.~\ref{model_predictions}(a)), although there still exists a jump in $\sigma$; this regime signifies CST. Eventually, when $p_0>p_0^*$, the system exhibits a Newtonian behavior. 

The parameter $\sigma^*$ distinguishes the liquid-like and solid-like behaviors. Although the model predicts DST for a constant $\sigma^*$ (Fig.~\ref{model_predictions}(a)), for the tissue models, $\sigma^*$ should not be constant. On the one hand, the system remains solid-like for $p_0<p_0^m$ due to the jamming transition, irrespective of the stress value. On the other hand, the system is always liquid-like when $p_0>p_0^*$ due to the transition at $p_0^*$. This implies that $\sigma^*$ becomes zero below $p_0^m$ and goes to $\infty$~\cite{footnote1} as $p_0>p_0^*$; therefore, it must depend on $p_0$. We obtain the precise $p_0$-dependence using our simulation data: we fit the flow curve, Eq.~(\ref{DSTmodel}), to the stress-controlled results, treating $\delta$ and $\sigma^*$ as fitting parameters. We find that a constant $\delta=0.6$ gives a good description of the data; however, consistent with our expectation, $\sigma^*$ depends on $p_0$, as shown in Fig.~\ref{model_predictions}(b). The line shows a power law fit: $\sigma^*=0.107 (p_0-3.823)^{3/2}$. Using this $\sigma^*$ and $\delta=0.6$, Fig.~\ref{model_predictions}(c) shows the comparison of the theoretical flow curves with those from the simulations, both for the rate-controlled and stress-controlled simulation data. Figure~\ref{model_predictions}(d) shows the behaviors of the model flow curves with a broad range of $p_0$ values. Similar to the constant $\sigma^*$ case (Fig.~\ref{model_predictions}(a)), the flow curves again show DST, followed by CST, and then near-Newtonian behaviors as $p_0$ increases. Despite this qualitative similarity, the quantitative natures of the flow curves in Figs.~\ref{model_predictions}(d) and \ref{model_predictions}(a) are different. This difference is crucial for a detailed comparison with the simulation data.

\begin{figure*}[t]
\centering
\includegraphics[width=17cm]{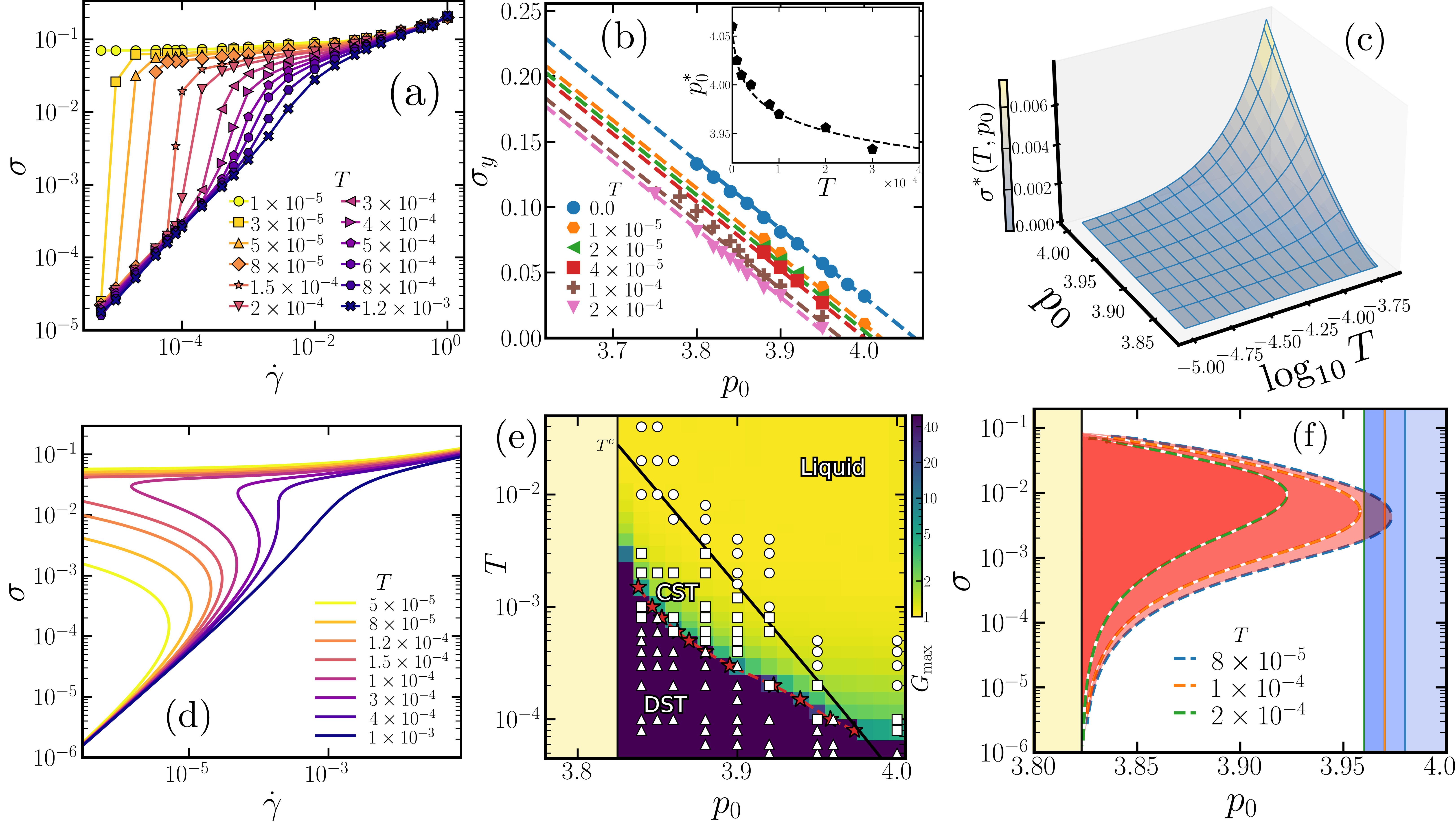}
	\caption{Simulation and theoretical predictions with varying $T$. (a) Simulation flow curves for $p_0 = 3.9$ and varying temperatures $T$. (b) $\sigma_y$ as a function of $p_0$ for various $T$; the inset shows the variation of $p_0^*$ with $T$, symbols are simulation data and dashed line is the fit of Eq.~\eqref{eq:pstarT}. (c) Characteristic stress for liquid-like to solid-like transition, $\sigma^*(p_0, T)$, plotted as a function of $p_0$ and temperature $T$ (see text for the details). (d) Theoretical flow curves for $p_0 = 3.88$ at different temperatures $T$. (e) Theoretical phase diagram in the $T$ vs. $p_0$ plane, where the color scale represents $G_{\max} = \mathrm{d}\log\sigma / \mathrm{d}\log\dot{\gamma}|_{\text{max}}$. We have also marked the different regimes. The dashed and solid lines mark the DST-CST and CST-Newtonian behavior boundaries, respectively. The symbols represent the simulation results: triangles indicate DST, squares indicate CST, and circles indicate Newtonian behavior. (f) $p_0^*$ and DST region decrease with increasing $T$. DST vanishes at a large enough $T$ when $p_0^*\leq p_0^m$.
	}
	\label{model_flow}
\end{figure*}

We now construct the analytical phase diagram in the $(\sigma,p_0)$ plane, which summarizes the different rheological regimes. We start with the constitutive equation, $\sigma = g(\sigma, \dot\gamma)$, and obtain the differential change in $\gdot$ with respect to $\sigma$ as,
$\frac{\mathrm{d}\dot\gamma}{\mathrm{d} \sigma} = (1-\frac{\partial g}{\partial \sigma})/\frac{\partial g}{\partial \dot\gamma}$. 
The DST boundary is given by $\frac{d\dot\gamma}{d \sigma} \to 0$ leading to $\frac{\partial g}{\partial \sigma} = 1$. Thus, the explicit DST condition becomes
\begin{equation}\label{DSTcondition}
	(1/\beta) \sigma^{-1/\beta-1} + f^{\prime}(\sigma)\,(\sigma_s^{-1/\beta}-\sigma_l^{-1/\beta})=0.
\end{equation}
We determine the DST boundary by simultaneously solving the constitutive equation, Eq.~(\ref{DSTmodel}), together with Eq.~(\ref{DSTcondition}) at each $p_0$. The line in Fig.~\ref{model_predictions}(e) gives the theoretical phase diagram. The shaded region inside this line gives DST, followed by CST outside it. The shear-thickening behavior disappears for $p_0>p_0^*$, at which the system exhibits Newtonian behavior. The symbols in Fig.~\ref{model_predictions}(e) show the simulation data in the regime where we could characterize DST in our simulations; the theory agrees well with the data. The jump in stress, $\Delta\sigma$, across the liquid-solid transition in the DST regime vanishes at a $p_0$, demarcating DST from CST. Figure~\ref{model_predictions}(f) shows the theoretical prediction (dashed line) for the vanishing of $\Delta \sigma$ with increasing $p_0$, and the symbols show the simulation data. The theoretical predictions again agree well with the simulation data.

We next investigate the $T$-dependence of the shear-thickening properties; this requires an enormous simulation effort to study rheological behavior across the entire $(p_0,T)$ plane. We first study the flow curves at a fixed $p_0$ with varying $T$, as shown in Fig.~\ref{model_flow}(a) for $p_0=3.90$ and different values of $T$. The SM Fig. S10 shows many such flow curves with varying $p_0$ and $T$. At $p_0=3.90$, the system shows DST at low $T$; however, as $T$ increases, the shear thickening becomes continuous (CST), and eventually the system behaves like a Newtonian fluid at large $T$. $T$ affects both $p_0^*$ and $\sigma^*$ that primarily govern the rheological properties. We first obtain the $T$-dependence of $p_0^*$. At a fixed $T$, we study the flow curves in the solid-like regime for various $p_0$ and, by analyzing the data using the Herschel-Bulkley form discussed in Sec.~\ref{noST_zeroT}, we obtain $\sigma_y(p_0,T)$. We then vary $T$ and obtain $\sigma_y(p_0,T)$ for different values of $T$. Figure~\ref{model_flow}(b) shows that $\sigma_y(p_0,T)$ has a nearly linear behavior as a function of $p_0$ at different $T$ in the regime of investigation ($\sigma_y$ deviates from the linear form as $\sigma_y\to0$~\cite{huang2022}, we excluded this very small $\sigma_y$-regime from the analysis). We fit the data with the linear form $\sigma_y(p_0,T)=C[p_0^*(T)-p_0]$, treating both $C$ and $p_0^*$ as fitting parameters. We find that $C\simeq 0.52\pm 0.02$ and varies only slightly. We treat it as a constant; Fig.~\ref{model_flow}(b) shows the fits with the constant $C=0.52$ with the data at different $T$. The fits also give $p_0^*(T)$ [inset of Fig.~\ref{model_flow}(b)]. The $T$-dependence of $p_0^*$ is well described by the functional form
\begin{equation}
	p_0^*(T) = c \left[ 1 - a \log(1 + b T ) \right],
	\label{eq:pstarT}
\end{equation}
with fitting parameters $c = 4.06$, $b = 304667$, and $a = 0.0064$. We show the fit with this function in the inset of Fig.~\ref{model_flow}(b).

We now focus on the $p_0$- and $T$-dependence of $\sigma^*(p_0,T)$. Since $p_0$ and $T$ are two independent control parameters, we propose the following general form
\begin{equation}\label{p*behavior}
	\sigma^*(p_0,T) = \kappa \mathcal{F}_1(p_0) \mathcal{F}_2(T),
\end{equation}
where $\kappa$ is a constant, and we will obtain the functional forms for $\mathcal{F}_1$ and $\mathcal{F}_2$ using the simulation data. Fitting the flow curves data (Fig.~\ref{model_flow}(a) and SM Fig. S12) with Eq.~(\ref{DSTmodel}) using $\sigma^*$ as a fitting parameter, we find $\sigma^*\sim T^{3/2}$ (SM Fig. S7). This implies, $\mathcal{F}_2(T) = T^{3/2}$. On the other hand, we have already obtained $\mathcal{F}_1(p_0)$ for $T=2\times 10^{-4}$ (Fig.~\ref{model_predictions}(b)): $\mathcal{F}_1(p_0)=(p_0-3.823)^{3/2}$. Furthermore, using the proportionality constant for $\sigma^*$ at $T=2\times 10^{-4}$, we get $\kappa=1127^{3/2}$.
Using all these expressions, we obtain
\begin{equation}\label{sigmastareq}
	\sigma^*(p_0,T) = \kappa T^{3/2} (p_0 - p_0^m)^{3/2}.
\end{equation}
Figure~\ref{model_flow}(c) shows the surface plot of $\sigma^*(p_0,T)$ in the $(p_0,T)$ plane; $\sigma^*$ increases as both $p_0$ and $T$ increase. The shear thickening disappears when $\sigma^*$ increases above the solid-regime stress value. Thus, we expect that shear thickening will cease to exist in the $(p_0,T)$ plane beyond a critical point, as we show below.

Given these analytical expressions for $p_0^*$ and $\sigma^*(p_0,T)$, we are now in a position to theoretically describe the flow properties at any point in the $(p_0,T)$ plane. We show the typical flow curves for $p_0=3.88$ and different values of $T$ in Fig.~\ref{model_flow}(d). Similar to the simulation results, the theory also predicts DST at small $T$, followed by CST, and then Newtonian-like behaviors at large $T$. We show the phase diagram, including the phase boundaries, in Fig. \ref{model_flow}(e) for the flow behavior in the $(p_0,T)$ plane. To obtain this phase diagram, we proceed as follows. Theoretically, we obtain the DST-CST boundary using the procedure as described above (Fig. \ref{model_predictions}(e)). The red dashed line in Fig. \ref{model_flow}(e) shows this boundary. On the other hand, the CST-Newtonian behavior boundary is given by Eq. (\ref{p*behavior}), and shown by the black solid line in Fig. \ref{model_flow}(e). To further compare with the simulation data and the DST phase diagram of the active vertex model, we use the same criterion as in Ref.~\cite{hertaeg2024}. We define the maximum slope: $G_\text{max}=\mathrm{d}\log\sigma/\mathrm{d}\log\gdot|_{\text{max}}$ and color code the theoretical flow curves according to their $G_\text{max}$ [Fig. \ref{model_flow}(e)]. In addition, we plot the simulation data points according to the following convention: a triangle indicates a flow curve exhibiting DST, a square indicates CST, and a circle indicates Newtonian behavior. The simulation data are in good agreement with the theoretical phase diagram. This phase diagram in the $(p_0,T)$ plane shows that DST vanishes for $p_0<p_0^m$ at all $T$ as $p_0^m$ is the jamming point, which is independent of $T$. However, the shear thickening also vanishes at large enough $T$. This large-$T$ effect is the consequence of varying $p_0^*$ with $T$. Thus, the shear thickening behavior disappears above the critical temperature $T^c$, as marked in Fig.~\ref{model_flow}(e).

For a deeper understanding of how DST vanishes, it is helpful to look at the phase diagram in the $(p_0,\sigma)$ plane with varying $T$. Figure~\ref{model_flow}(f) shows the behaviors at three different $T$. We have already shown how $p_0^*(T)$ decreases with increasing $T$ in the inset of Fig.~\ref{model_flow}(b). The solid lines in Fig.~\ref{model_flow}(f) show the values of $p_0^*$ at these three $T$. The dashed lines are the theoretical phase diagrams at these values of $T$. The left-hand side boundary in Fig.~\ref{model_flow}(f) is given by $p_0^m$, the jamming point of the tissue. On the other hand, the boundary demarcating CST and the Newtonian behavior is controlled by $p_0^*$. However, the boundary between DST and CST is governed by both $\sigma^*(p_0,T)$ and $p_0^*$. As $T$ increases, the DST regime shrinks. Eventually, at large enough $T=T^c$, $p_0^*(T^c)$ will coincide with $p_0^m$ and shear thickening will disappear at this $T^c$. Below this $T^c$, as $p_0$ increases, the system follows DST, CST, and then Newtonian behavior.

\section{Discussions and Conclusions}
\label{disc}

To conclude, we have introduced a constitutive model to describe the DST in the vertex model of confluent biological tissues. The shape index $p_0$ in this model plays a crucial role in governing its properties. The model shows an unusual jamming transition at $p_0^m$, where the elastic modulus vanishes but $\sigma_y$ does not; the latter vanishes at a higher value of $p_0=p_0^*$~\cite{huang2022}. The system exhibits DST in this intermediate regime, $p_0^m < p_0 < p_0^*$. DST has been widely studied in dense granular suspensions, where it arises from a transition from frictionless to frictional rheology as the number of contacts between rough particles increases as the solvent, acting as a lubrication layer, drains out. By contrast, the mechanism in the tissue model is entirely different: the system switches from a liquid-like to a solid-like behavior. The crucial ingredients are the finite values of $\sigma_y$ and $\tau$ in the unjammed phase. When $\gdot$ is small, such that $\gdot^{-1}>\tau$, the system has enough time to relax the shear-induced deformation and behaves as a liquid. In the other regime, when $\gdot^{-1}<\tau$, the system cannot relax the shear-induced deformation, leading to finite stress and solid-like behavior. The nature of this liquid-like to solid-like transition is governed by the magnitude of $\sigma_y$: when $\sigma_y$ is large, the system shows DST. As $\sigma_y$ decreases with increasing $p_0$, we find CST and eventually, when $p_0>p_0^*$, where $\sigma_y=0$, the system behaves as a Newtonian fluid.

The discovery of DST in tissue models broadens the field of interesting rheological behaviors in physical systems. Existing particulate models showing DST require a solvent that first provides lubrication for frictionless rheology, then drains to form particle contacts and enable frictional rheology. The Wyart-Cates model brilliantly describes this generic, intuitive, yet powerful mechanism~\cite{wyart2014}. To the best of our knowledge, there is no single component model of DST in particulate systems. The tissue models are the first such instance, showing DST in a single-component system, albeit with a complex multi-point interaction energy (Eq.~\eqref{normalizedH}). Similar to the Wyart-Cates model, we have demonstrated that we can understand the DST even in this system in terms of fractions of solid-like cells, although the mechanism is entirely different. The presence of non-zero $\sigma_y$ in the unjammed phase of the tissue model plays a crucial role in governing the transition from liquid-like to solid-like branches. In a future study, we will address the detailed mechanism of how $\sigma_y$ governs this transition between the two rheological behaviors.

What is the role of fluctuations, and their specific natures, that govern DST in tissue models? In a recent work, Hertaeg {\it et al} have shown DST in the active vertex model~\cite{hertaeg2024}. On the other hand, in this work, we have studied DST in the thermal vertex model. Our theory for the DST shows that the system must have some form of fluctuation that allows a finite $\tau$ that competes with the shear time scale, $\gdot^{-1}$, in the unjammed phase with non-zero $\sigma_y$. The role of fluctuations is only to fluidize the tissue, and their precise nature is not crucial. In fact, both thermal noise and active self-propulsion fluidize the tissue monolayer~\cite{activematterreview,bi2016,sadhukhan2021}. Consistently, we see a very similar phase diagram for both the active vertex model~\cite{hertaeg2024} and the thermal system results presented here (Fig.~\ref{model_flow}(e)). Our results also demonstrate that DST does not result from the nontrivial nature of activity but is a property of the tissue model itself. An interesting future direction is to extend the theory to active systems with finite persistence time, and we expect the predictions of that theory to agree well with the simulation results of Ref.~\cite{hertaeg2024}. Furthermore, studying rheological properties in the presence of other forms of activity, such as fluctuations in cell volume~\cite{tjhung2017,li2025} and division-apoptosis~\cite{fernandez2017,czajkowski2019}, will provide further insight into the role of fluctuations in governing DST.

Three distinct properties of the system govern the boundaries of the shear-thickening behavior. The jamming point, $p_0^m$, gives the starting point of the shear thickening. For $p_0<p_0^m$, the system is solid with both $G$ and $\sigma_y$ being non-zero at $T=0$. On the other extreme, $p_0=p_0^*$, where $\sigma_y$ becomes zero, marks the end-point of shear thickening, and the system behaves as a Newtonian fluid beyond this point. In between, the boundary separating DST from CST is governed by both $\sigma^*$ and $p_0^*$. In addition, we find that $\sigma^*$ vanishes at $p_0^m$ (Eq.~\eqref{sigmastareq}), implying that the system is always solid-like and setting the lower boundary for shear thickening. This result is consistent with the fact that the unjamming point, which is independent of $T$, dictates the lower boundary. We will investigate the rheological properties in the vicinity of this boundary in a future work. On the other hand, with varying $T$, both $p_0^*$ and $\sigma^*$ change. $p_0^*$ decreases as $T$ increases and eventually becomes the same as $p_0^m$ marking the end point of shear thickening with increasing $T$. The phase diagram clearly shows that the glass line, which depends on $T$ or activity and vanishes at $p_0^m$, is irrelevant for the DST and the rheological properties.

As epithelial monolayers act as a protective barrier, DST has critical consequences for various physiological processes, as it enables rapid changes in material properties in response to external mechanical perturbations. In addition, cell shape, which governs several biological functions, directly dictates cellular-level stress and thus DST. With the advent of novel experimental techniques, it is now possible to study the rheological properties of a tissue monolayer~\cite{duque2024,yang2025}. Despite these advancements, studying DST will be nontrivial, as $p_0$ is not directly measurable. However, with proper biochemical design, it should be possible to tune the system to a regime such that the condition $p_0^m<p_0<p_0^*$ is satisfied. In addition, since the activity or effective temperature should not be too high to wash away the jamming properties, targeting junctional molecules without increasing cellular activity might be an appropriate approach. Our constitutive model yields several key predictions for the rheology of epithelial tissues. The vanishing of stress jump at a $p_0$ separating DST from CST (Fig.~\ref{model_predictions}(f)), the DST phase diagram both in $(\sigma,p_0)$ (Fig.~\ref{model_predictions}(e)) and $(T,p_0)$ plane (Fig.~\ref{model_flow}b), and the S-shaped flow curves (Fig.~\ref{model_predictions}(c)) agree well with our simulation data. We believe that our theory will be instrumental in testing DST properties in biological tissues and understanding its deeper implications.

\section*{Acknowledgments}
We acknowledge the support of the Department of Atomic Energy, Government of India, under Project identification No. RTI 4007.

\bibliography{ref.bib}

\clearpage
\onecolumngrid


\section*{Supplementary material}

# Supplementary Material: "A constitutive model for discontinuous shear thickening in epithelial tissues"

Tanmoy Ghosh, Kabir Ramola, and Saroj Kumar Nandi
*Tata Institute of Fundamental Research, Hyderabad - 500046, India*

In this supplementary material, we provide further details of the simulations and the definitions of the self-overlap function that characterizes relaxation dynamics, the area, perimeter, center of mass, and the aspect ratio of a cell. We then show the macroscopic stress distribution from our simulations. We next provide details on how we obtained the results of the stress-controlled simulations, their comparison with the rate-controlled data, and how these results depend on the system sizes. To characterize the static aspects, we also show the shape-index distribution. We then present the theoretical fits with the simulation data to obtain the parameters of the theory as well as the temperature-dependence of $\sigma^*$. Next, we present the stress-strain curves with varying strain rates. We have then showed how the theoretical flow curves depend on the exponent $\beta$. Finally, we present the flow curves for various $p_0$ and $T$.

## S1. Shear protocol and boundary conditions

As detailed in the main text, we have studied the rheological properties of the vertex model under an external shear, both at zero and non-zero temperature $T$. We impose shear in vertex-model simulations by replacing the standard periodic boundary conditions with the Lees–Edwards boundary conditions [1, 2], which account for the relative motion of periodic images under shear. In this framework, neighboring periodic replicas of the simulation domain are displaced laterally by a shear-dependent offset, as shown in Fig. S1, generating a uniform shear deformation across the system. Vertices crossing the boundary in the flow direction ($x$) are remapped using standard periodic rules. For crossings in the gradient direction ($y$), the $x$-coordinate of the vertex is first shifted by an amount $\Delta x = \gamma L_y$ before applying periodicity.

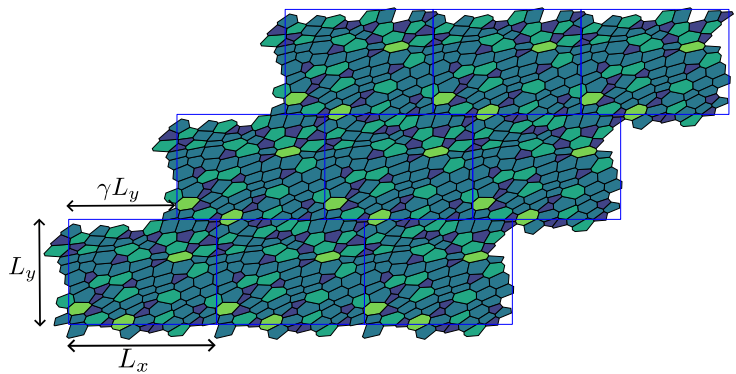

FIG. S1. Schematic illustration of the Lees–Edwards boundary condition used in the simulation of the vertex model. Periodic replicas of the simulation box, shifted by a strain-dependent displacement in the flow direction, produce a uniform shear deformation while preserving periodicity.

## S2. Self overlap function to quantify dynamics

We quantify the relaxation dynamics of the system using the overlap function $Q(t)$, defined as

$$Q(t) = \left\langle \frac{1}{N} \sum_{i=1}^{N} W[a - |\mathbf{r}_i(t) - \mathbf{r}_i(0)|] \right\rangle, \tag{S1}$$

where $\mathbf{r}_i(t)$ is the center-of-mass position of cell $i$ at time $t$, $N$ is the total number of cells. The angular brackets $\langle \cdots \rangle$ denote an average over different time origins $t$ and over many ensembles of the system. The window function $W(x)$

is chosen as the Heaviside step function,

$$W(x) = \begin{cases} 1, & x > 0, \\ 0, & x \leq 0. \end{cases} \tag{S2}$$

Consistent with earlier studies, we have chosen the cutoff parameter $a = 0.3$ [3, 4]. The relaxation time $\tau_\alpha$ is defined when $Q(t)$ becomes $1/3$, explicitly, $Q(\tau = \tau_\alpha) = 1/3$.

### S3. Area, perimeter, and center of mass of a polygon

As described in the main text, cells are represented as polygons within the vertex model. We compute the center of mass from the positions of the vertices of a cell [3]. Let us denote the cell vertices as $\{(x_i, y_i)\}_{i=1}^n$, where $n$ denotes the total number of vertices of the cell. Without loss of generality, we can consider that the vertices are ordered, either clockwise or counterclockwise, along the cell boundary. Then we obtain the center of mass of the cell as

$$x_c = \frac{1}{6A} \sum_{i=1}^n (x_i + x_{i+1})(x_i y_{i+1} - x_{i+1} y_i), \tag{S3}$$

$$y_c = \frac{1}{6A} \sum_{i=1}^n (y_i + y_{i+1})(x_i y_{i+1} - x_{i+1} y_i). \tag{S4}$$

The cell area $A$ is computed using the shoelace formula,

$$A = \frac{1}{2} \sum_{i=1}^n (x_i y_{i+1} - x_{i+1} y_i). \tag{S5}$$

The cell perimeter $P$ is obtained as the sum of the lengths of the edges that connect successive vertices;

$$P = \sum_{i=1}^n \sqrt{(x_{i+1} - x_i)^2 + (y_{i+1} - y_i)^2}, \tag{S6}$$

where we have assumed the periodicity of the indices as the cell boundary forms a closed loop:

$$x_{n+1} = x_1, \qquad y_{n+1} = y_1. \tag{S7}$$

### S4. Aspect ratio from the moment of inertia tensor

To characterize the cell-shape anisotropy, we computed the aspect ratio from the eigenvalues of the moment of inertia tensor [5, 6]. Let us consider that the center of mass of the cell is given by $(x_c, y_c)$. We first shift the vertex coordinates to the center of mass reference frame,

$$\tilde{x}_i = x_i - x_c, \qquad \tilde{y}_i = y_i - y_c. \tag{S8}$$

We then calculate the components of the moment of inertia tensor in this reference frame using standard polygon expressions,

$$I_{xx} = \frac{1}{12} \sum_{i=1}^n (\tilde{x}_i \tilde{y}_{i+1} - \tilde{x}_{i+1} \tilde{y}_i)(\tilde{y}_i^2 + \tilde{y}_i \tilde{y}_{i+1} + \tilde{y}_{i+1}^2), \tag{S9}$$

$$I_{yy} = \frac{1}{12} \sum_{i=1}^n (\tilde{x}_i \tilde{y}_{i+1} - \tilde{x}_{i+1} \tilde{y}_i)(\tilde{x}_i^2 + \tilde{x}_i \tilde{x}_{i+1} + \tilde{x}_{i+1}^2), \tag{S10}$$

$$I_{xy} = \frac{1}{24} \sum_{i=1}^n (\tilde{x}_i \tilde{y}_{i+1} - \tilde{x}_{i+1} \tilde{y}_i)(\tilde{x}_i \tilde{y}_{i+1} + 2\tilde{x}_i \tilde{y}_i + 2\tilde{x}_{i+1} \tilde{y}_{i+1} + \tilde{x}_{i+1} \tilde{y}_i), \tag{S11}$$

with periodic indexing such that $\tilde{x}_{n+1} = \tilde{x}_1$ and $\tilde{y}_{n+1} = \tilde{y}_1$.

We then obtain the moment of inertia tensor as

$$\mathbf{I} = \begin{pmatrix} I_{xx} & I_{xy} \\ I_{xy} & I_{yy} \end{pmatrix}. \tag{S12}$$

Diagonalization of $\mathbf{I}$ yields two non-negative eigenvalues $\lambda_1 \geq \lambda_2$, corresponding to the principal moments of inertia. The cell aspect ratio (AR) is defined as

$$\text{AR} = \sqrt{\frac{\lambda_1}{\lambda_2}}, \tag{S13}$$

with AR = 1 indicating an isotropic cell.

The orientation of the cell's major axis is given by the eigenvector associated with $\lambda_1$. If $\mathbf{v} = (v_x, v_y)$ denotes this eigenvector, the major-axis angle is computed as

$$\theta = \arctan\left(\frac{v_y}{v_x}\right), \tag{S14}$$

and is reported modulo 180°.

### S5. Macroscopic stress distribution

We study the probability distribution $P(\sigma)$ of the macroscopic shear stress time series for different imposed strain rates $\dot{\gamma}$, while keeping the target shape index fixed at a particular value of $p_0$. Figure S2(a) shows $P(\sigma)$ for $p_0 = 3.88$. At low strain rates ($1 \times 10^{-5}$ and $4 \times 10^{-5}$) and at high strain rates ($1 \times 10^{-4}$ and $2 \times 10^{-4}$), the stress distribution remains unimodal, indicating relatively homogeneous stress fluctuations. By contrast, at intermediate strain rates ($6 \times 10^{-5}$ and $8 \times 10^{-5}$), corresponding to the discontinuous shear thickening (DST) regime, the stress distribution becomes distinctly bimodal. This bimodality reflects the coexistence of low and high-stress states and is a characteristic signature of DST observed in dense suspensions [7, 8]. The instantaneous shear time series also reflects this unimodal vs bimodal nature of $\sigma$ at varying shear rates [Fig. S2(b)].

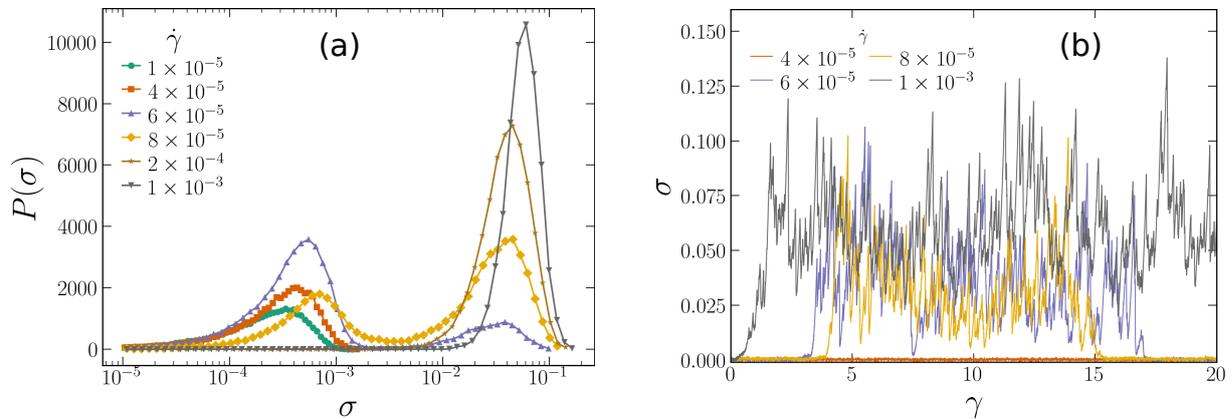

FIG. S2. (a) Probability distributions $P(\sigma)$ of the macroscopic shear stress for a range of imposed strain rates $\dot{\gamma}$. (b) Time series of the macroscopic stress $\sigma$ as a function of strain $\gamma$ for various strain rates, as indicated in the legend. All data correspond to $p_0 = 3.88$ and system size $N = 100$.

### S6. Shape index distribution

Within the vertex model, the observed shape index, $q_i = P_i/\sqrt{A_i}$, plays a crucial role governing the static properties of the system. This geometric observable has been widely used in the literature to distinguish between the solid-like and the liquid-like states of the system [9, 10]. This distribution reveals the emergence of heterogeneity and the coexistence of solid-like and liquid-like regions during the DST transition. We have analyzed the shape index $q_i$ for

each cell of the model and obtained the distribution $P(q_i)$ from the configurations drawn at the steady state for various values of the $\dot{\gamma}$, shown in Fig S3(a). We also show the distribution of the normalized perimeter stiffness $p_i = P_i/\sqrt{A_{0,i}}$ in Fig. S3(b).

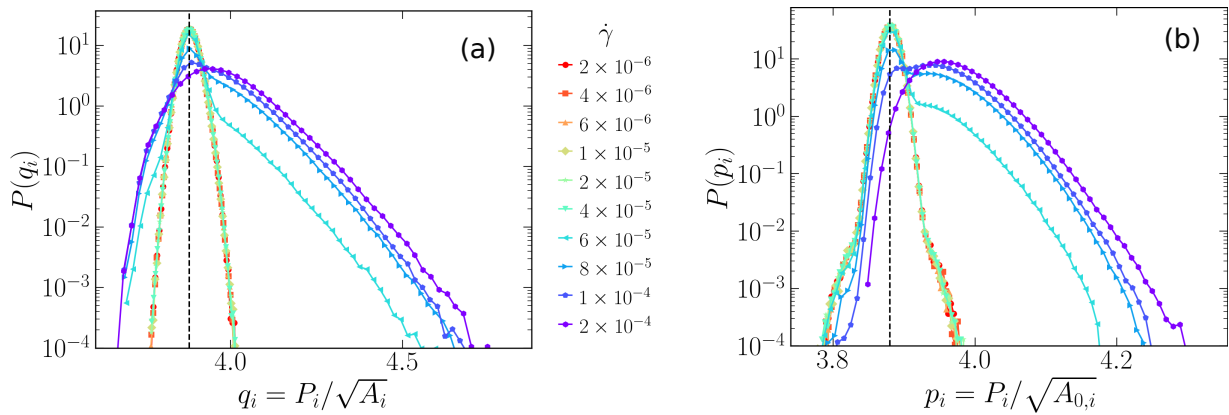

FIG. S3. Distribution of the structural parameters. (a) Probability distributions of the shape index $q_i = P_i/\sqrt{A_i}$ and (b) that of the normalized perimeter stiffness $p_i = P_i/\sqrt{A_{0,i}}$ for different imposed shear rates $\dot{\gamma}$, ranging from $2 \times 10^{-6}$ to $2 \times 10^{-4}$.

### S7. Stress controlled simulations

Figure S4 illustrates the system response under stress-controlled driving and the emergence of S-shaped flow curves. In these simulations, a constant macroscopic shear stress $\sigma_0$ is imposed and the resulting strain-rate response $\dot{\gamma}(t)$ is monitored [11]. Figures S4(a) and (b) show the time evolution of the accumulated strain $\gamma(t)$ and the corresponding instantaneous strain rate $\dot{\gamma}(t)$, respectively, for $p_0 = 3.9$ and different imposed stresses $\sigma_0$. To construct the flow curve, we consider only the steady-state portion of the strain rate, discarding the transient regime. We show the comparison of the steady-state flow curves obtained under rate-controlled and stress-controlled protocols for $p_0 = 3.88$ and $p_0 = 3.9$ in Fig. S4(c). While rate-controlled simulations yield monotonic flow curves, stress-controlled simulations reveal an S-shaped relationship between stress and strain rate. The negative-slope branch of this curve is dynamically inaccessible under a rate-controlled scenario, consistent with shear-thickening instabilities.

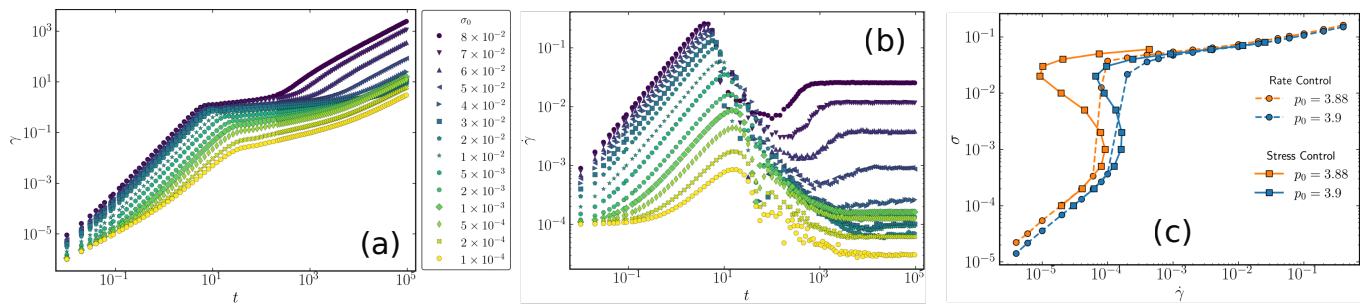

FIG. S4. Obtaining the flow curves from the stress-controlled simulations. (a) Time evolution of the accumulated strain $\gamma(t)$ for different imposed stresses $\sigma_0$ at $p_0 = 3.9$. (b) The instantaneous strain rates, $\dot{\gamma}(t)$, corresponding to the data shown in (a). (c) Steady-state values of $\dot{\gamma}$ at a particular stress gives the flow curves $\sigma(\dot{\gamma})$. We show the comparison of the stress-controlled curves (solid lines) with the rate-controlled data (dashed lines) for $p_0 = 3.88$ and $p_0 = 3.9$.

### S8. System-size dependence of the flow curves

Figure S5 compares the rheological response of the system under rate-controlled and stress-controlled driving for different system sizes. In the rate-controlled protocol Fig S5(a), the flow curves collapse reasonably well across system sizes, exhibiting no significant system-size dependence. By contrast, the stress-controlled protocol [Fig S5(b)] exhibits

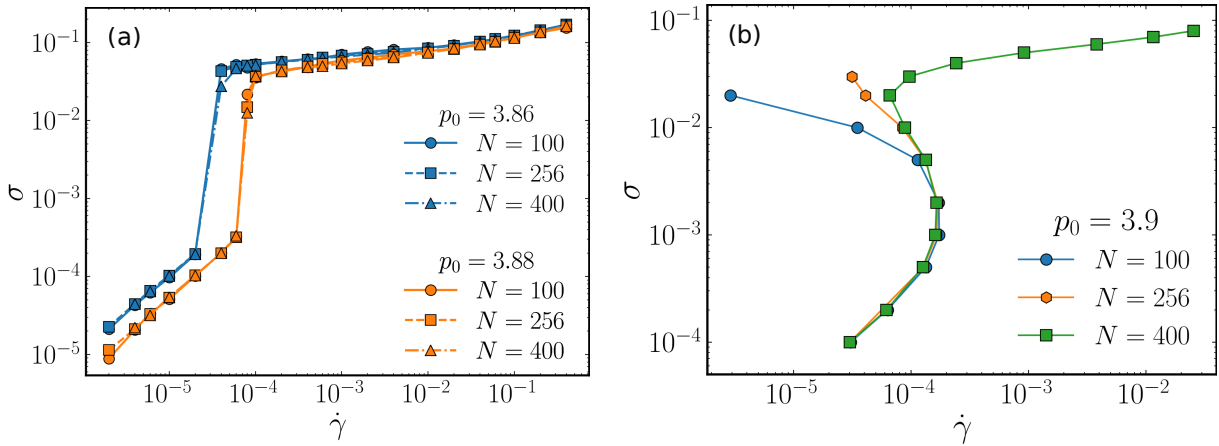

FIG. S5. System-size dependence of the flow curves. (a) Rate-controlled flow curves for two target shape indices, $p_0 = 3.86$ and $3.88$, shown for three system sizes $N = 100$, $256$, and $400$. (b) Stress-controlled flow curves at $p_0 = 3.9$ for the same system sizes.

a non-monotonic, S-shaped flow curve whose extent depends on system size. As the system size increases, the S-shaped region becomes progressively less pronounced, indicating that the non-monotonic behavior weakens with increasing $N$ and suggesting that it may vanish in the thermodynamic limit. Similar results have been presented elsewhere for the DST in dense particulate suspensions [12, 13].

### S9. Individual fits for liquid-like and solid-like regimes of the flow curves

As discussed in the main text, we have fitted the liquid-like and solid-like regimes of the flow curves separately to obtain various model parameters.

The solid-like branch shows a Herschel-Bulkley yield-stress material behavior [14–16],

$$\sigma_s = \sigma_y + B\dot\gamma^n = C\left(p_0^* - p_0\right) + B\dot\gamma^n, \quad (S15)$$

where $p_0^*$ denotes the transition to the Newtonian behavior, marked by the vanishing $\sigma_y$. At $T = 2 \times 10^{-4}$, we obtain

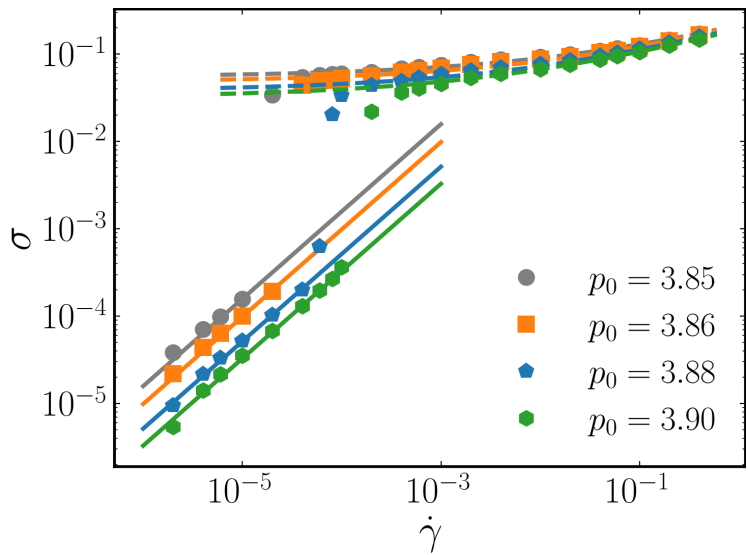

FIG. S6. Fitting of Eqs. (S15) and (S16) with solid-like and liquid-like regimes at $T = 2 \times 10^{-4}$. The symbols represent the rate-controlled flow curves for $p_0 = 3.85$, $p_0 = 3.86$, $p_0 = 3.88$, and $p_0 = 3.9$; the dashed lines show the fits with the solid-like branch using a Herschel-Bulkley form (Eq. S15), while solid lines correspond to fits with the liquid-like branch (Eq. S16).

$p_0^* = 3.96$ (Fig. 6(b) in the main text).

We fit the liquid branch using the following form,

$$\sigma_\ell = A \dot{\gamma} (p_0 - p_0^m)^{-\alpha}, \tag{S16}$$

where $p_0^m$ is the rigidity point.

We can fit Eqs. (S15) and (S16) with two sets of data, one for the solid-like regime and another for the liquid-like regime, respectively, and obtain the values for the constants as follows:

$$\alpha = 1.5, \quad A = 0.07, \quad C = 0.52, \quad B = 0.16, \quad n = 0.33. \tag{S17}$$

Figure S6 shows the comparison of the analytical forms, Eqs. (S15) and (S16), with the simulation data for various $p_0$ values using the same set of parameters as given above. The excellent agreement shows the applicability of the scaling forms for the individual regimes.

### S10. Temperature dependence of $\sigma^*$

We now present the analysis of the temperature dependence of the characteristic stress scale $\sigma^*$. As discussed in the main text, we have proposed a separation of variables as the $p_0$ and $T$ are independent control parameters. Thus, we have written

$$\sigma^*(p_0, T) = \kappa F_1(p_0) F_2(T), \tag{S18}$$

where $\kappa$ denotes a constant, while $F_1$ and $F_2$ depend solely on $p_0$ and $T$, respectively. Fitting the simulation data for different values of $p_0$ yields

$$\sigma^* \sim T^{3/2}, \tag{S19}$$

implying

$$F_2(T) \sim T^{3/2}. \tag{S20}$$

Figure S7 shows the comparison of the fits (solid lines) with the simulation data (symbols). The excellent agreement of the data with the power-law trends with the identical slopes confirms that the temperature dependence of $\sigma^*$ is well-described by a $T^{3/2}$ scaling, independent of $p_0$.

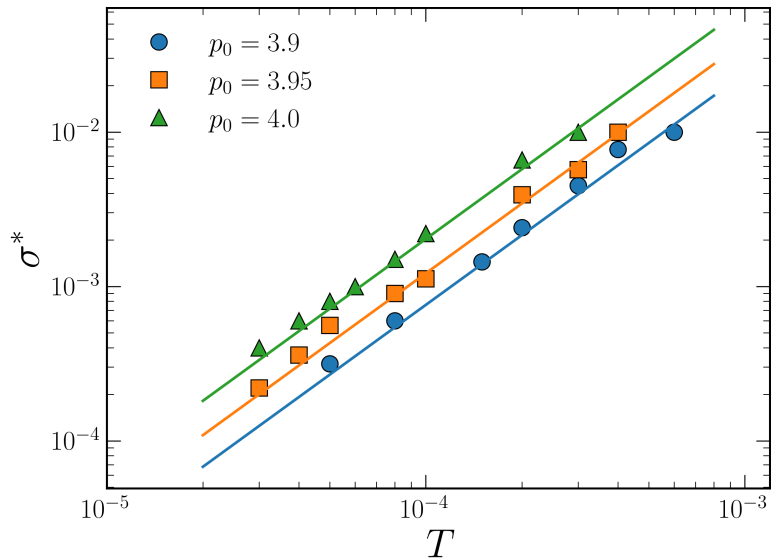

FIG. S7. Temperature dependence of the characteristic stress $\sigma^*$ for $p_0 = 3.9$ (blue circles), $p_0 = 3.95$ (orange squares), and $p_0 = 4.0$ (green triangles). Symbols denote simulation data, while solid lines indicate power-law fits with slope 3/2. The common scaling exponent implies a similar $T$-dependence at varying $p_0$.

### S11. Stress-strain curves for varying shear rates $\dot\gamma$

We show the stress - strain curves at a fixed shape parameter $p_0 = 3.9$ for different shear rates $\dot\gamma$ in Fig. S8. Starting from an undeformed configuration, the stress $\sigma$ initially increases with strain $\gamma$, reaching a pronounced overshoot before relaxing toward a steady-state value. The magnitude of the stress overshoot increases with increasing shear rate, while the steady-state stress attained at large strain shows a weaker $\dot\gamma$-dependence.

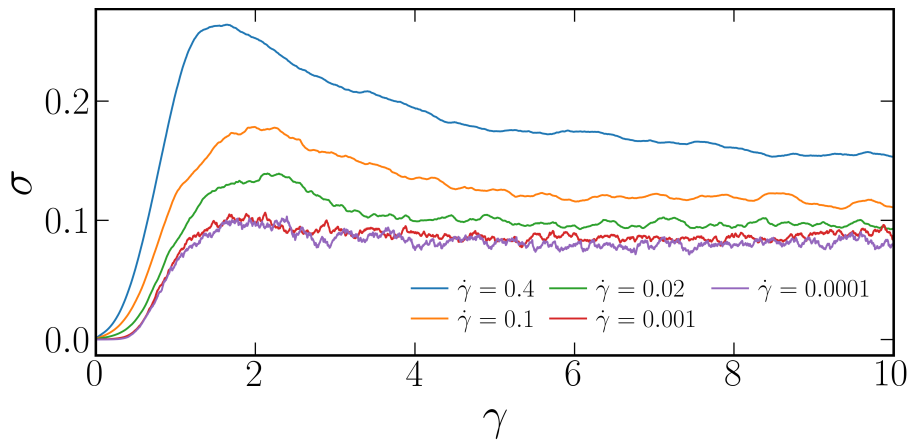

FIG. S8. Stress $\sigma$ as a function of accumulated strain $\gamma$ for $p_0 = 3.9$ at different imposed shear rates $\dot\gamma$. The curves show an initial elastic-like increase, followed by a stress overshoot and subsequent relaxation to a steady-state value. The height of the overshoot increases with increasing shear rate, while the steady-state stress exhibits a weaker rate dependence.

### S12. The theoretical flow curves at varying $\beta$ and $f(\sigma)$

As mentioned in the main text, the existence of DST in our theory does not require a specific value of $\beta$ or a particular form of $f(\sigma)$. Although the quantitative nature of the curves depends on $\beta$, the qualitative S-shaped nature of the flow curves, Eq. (7) in the main text, remains intact with varying $\beta$. As shown in Fig. S9(a) for various values of $\beta$, the width of the S-shape increases as $\beta$ decreases; however, the curves still show the S-shaped characteristic.

On the other hand, we also consider several forms of $f(\sigma)$, as shown in Fig. S9(b), they all have sigmoidal nature, that is, $f(\sigma) = 0$ at small $\sigma$ and $f(\sigma) = 1$ at large $\sigma$. The quantitative nature of the flow curves changes [Fig. S9(c)], however, the S-shape still remains, signifying the presence of DST.

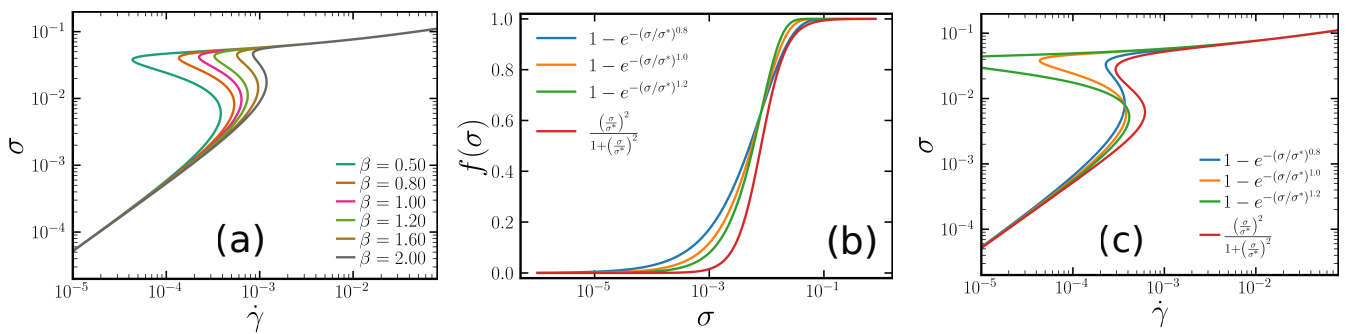

FIG. S9. (a) Theoretical flow curves $\sigma(\dot\gamma)$ computed for different values of the exponent $\beta$ [see Eq. (7) in the main text]. Although the quantitative nature of the curves depends on the value of $\beta$, the characteristic S-shaped, nonmonotonic form remains preserved for all $\beta$ considered, demonstrating that the flow-curve topology is insensitive to the specific choice of $\beta$. (b) Different sigmoidal forms of $f(\sigma)$ satisfying $f(\sigma) \to 0$ at small $\sigma$ and $f(\sigma) \to 1$ at large $\sigma$. (c) Corresponding flow curves obtained using the different forms of $f(\sigma)$. Although the detailed shape of the curves changes, the S-shaped topology persists, demonstrating that the emergence of discontinuous shear thickening (DST) is insensitive to the specific choice of $\beta$ or the functional form of $f(\sigma)$.

## S13. Rate-controlled flow curves at different $p_0$ and $T$

For the phase diagram in the $(p_0 - T)$ plane, we must investigate the nature of the flow curves over the entire plane; this requires a substantial amount of simulation effort, generating data at many values of $p_0$ and $T$. We have used the rate-controlled simulations for this purpose. The data allow us to obtain the theory for both varying $p_0$ and $T$. Figure S10 shows the flow curves, $\sigma(\dot{\gamma})$, obtained for various $p_0$ and $T$ at a fixed system size $N = 100$. For each value of $p_0$, different flow curves correspond to various temperatures, as shown by the color bar next to each of the flow curves.

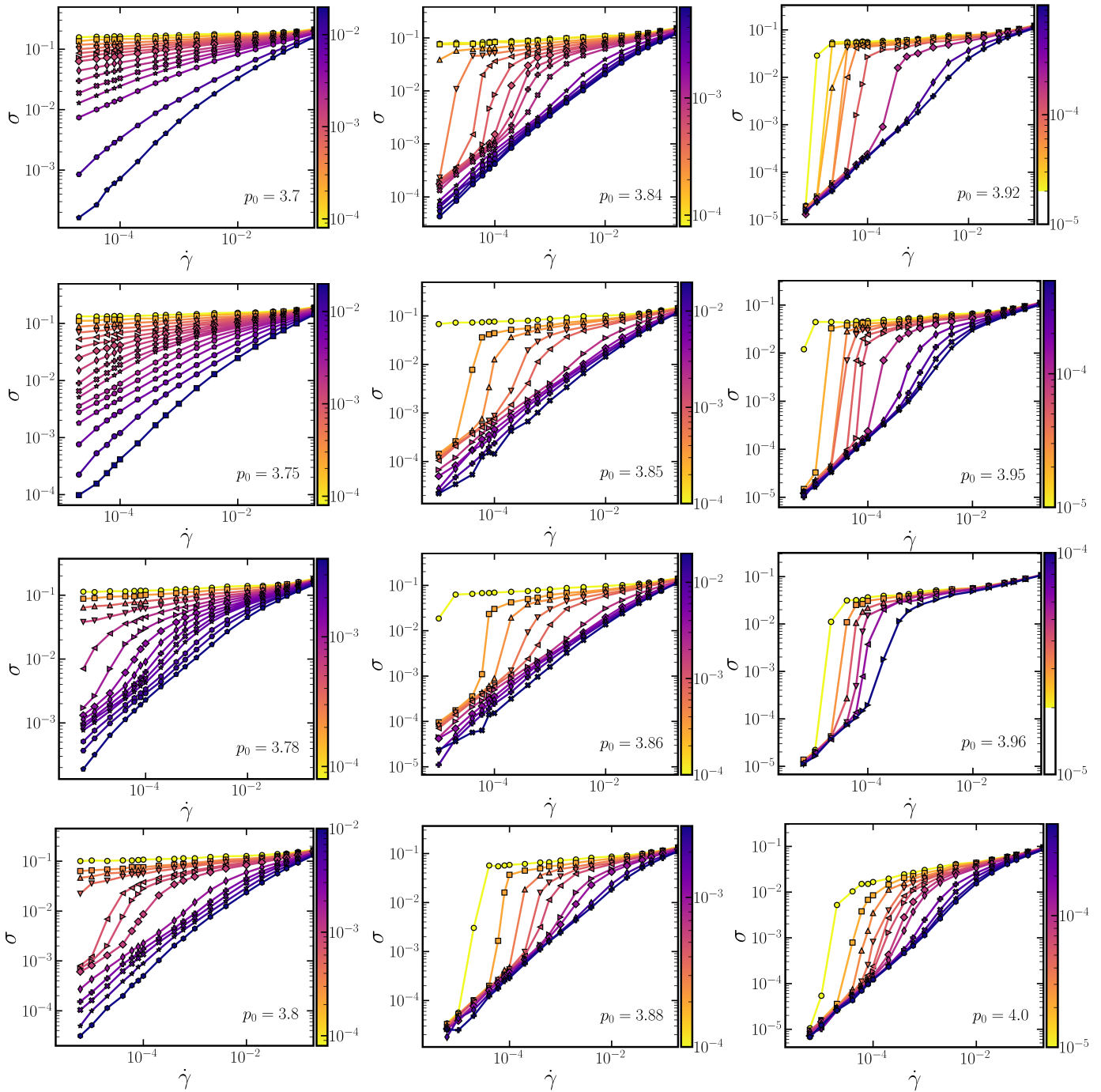

FIG. S10. Rate-controlled flow curves $\sigma(\dot{\gamma})$ for different values of the shape parameter $p_0$ and temperature $T$ at fixed system size $N = 100$. Each panel corresponds to a different $p_0$, as indicated. Color-bars denote temperature values in a logarithmic scale, with darker colors representing higher $T$ and lighter colors lower $T$.



\end{document}